
%
%
\documentstyle{amsppt}
\TagsOnRight
\catcode`\@=11
\def\logo@{}
\catcode`\@=13
\parindent=8 mm
\magnification 1200
\hsize = 12.2  cm
\vsize = 19.4  cm
\hoffset = 1 cm
\parskip=\medskipamount
\baselineskip=14pt

\def \smaller {\eightpoint}
\def \wt {\widetilde}
\def \wh {\widehat}

\def \ra {\rightarrow}
\def \hra {\hookrightarrow}
\def \lra {\longrightarrow}
\def \lmt {\longmapsto}
\def \a {\alpha}

\def \d {\delta}
\def \e {\epsilon}
\def \g {\gamma}
\def \G {\Gamma}
\def \k {\kappa}

\def \l {\lambda}

\def \m  {\mu}

\def \r {\rho}
\def \s {\sigma}

\def \th {\theta}
\def \t {\tau}
\def \o {\omega}

\def \z {\zeta}

\def \ss {\subset}
\def \sps {\supset}
\def \Lg {\widetilde{\frak g}}

\def \LGp {\widetilde{\frak G}^+}

\def \Lgp {\widetilde{\frak g}^+}
\def \Lgm {\widetilde{\frak g}^-}
\def \LgA {\widetilde{\frak g}_A}
\def \lgp {\widetilde{\frak g}_+}
\def \lgm {\widetilde{\frak g}_-}
\def \AB {\bold{A}}

\def \CC {\Cal C}
\def \DD {\Cal D}

\def \II {\Cal I}
\def \JJ {\Cal J}

\def \LL {\Cal L}
\def \MM {\Cal M}
\def \NN {\Cal N}
\def \LL {\Cal L}
\def \OO {\Cal O}
\def \PP {\Cal P}
\def \QQ  {\Cal Q}
\def \SS {\Cal S}
\def \TT {\Cal T}
\def \di {\partial}

\def \ln{\text{ln}}
\def\Bl{\phantom{.}}
\hyphenation{Darboux Liouville Hurtubise Harnad Adams Arnold Helminck}
\font\titlefont=cmbx10 scaled\magstep2
\font\sectionfont=cmbx10 at 14pt
\font\subsectionfont=cmbx12
\font\subsubsectionfont=cmbx10
\vskip 1 cm
\noindent{\titlefont
Isospectral Flow and Liouville-Arnold  \newline\noindent
Integration in Loop Algebras}${}^\dag$
\rightheadtext{}
\leftheadtext{}
\footnote""
{\hskip -.8 cm ${}^\dag$Lectures presented at the VIIIth Scheveningen
Conference: {\it Algebraic Geometric Methods in Mathematical Physics}, held at
Wassenaar, the Netherlands, Aug\. 16-21, 1992. To appear in: Springer Lecture
Notes in Physics (1993). Research supported in part by the Natural Sciences and
Engineering Research Council of Canada and the Fonds FCAR du Qu\'ebec.}
\footnote""{\hskip -.8 cm $^1$e-mail address:
harnad\@alcor.concordia.ca  {\it \ or\ }  harnad\@mathcn.umontreal.ca}
\bigskip \bigskip \noindent J.~Harnad${}^1$
\medskip \noindent
{\eightpoint   Department of Mathematics and Statistics, Concordia
University
\newline \noindent
 7141 Sherbrooke W., Montr\'eal, Canada H4B 1R6, \ and \newline \noindent
 Centre de recherches math\'ematiques, Universit\'e de Montr\'eal
\newline \noindent
  C.~P. 6128-A, Montr\'eal, Canada H3C 3J7}
\bigskip \bigskip
\noindent {\bf Abstract.}
 Some standard examples of Hamiltonian systems that are integrable by
classical means are cast within the framework of isospectral flows in loop
algebras. These include: the Neumann oscillator, the cubically nonlinear
Schr\"odinger systems and the sine-Gordon equation. Each system has an
associated invariant spectral curve and may be integrated via the
 Liouville-Arnold technique. The linearizing map is the Abel map to the
associated Jacobi variety, which is deduced through separation of variables
in hyperellipsoidal coordinates. More generally,  a family of moment maps is
 derived, embedding certain finite dimensional symplectic manifolds, which
arise through Hamiltonian reduction of symplectic vector spaces, into rational
 coadjoint orbits of loop algebras $\wt{\frak{g}}^+\ss\wt{\frak{gl}}(r)^{+}$.
 Integrable Hamiltonians are obtained by restriction of elements of the ring of
 spectral invariants to the image of these moment maps; the isospectral
property follows from the Adler-Kostant-Symes theorem.  The structure of the
generic spectral curves   arising through the moment map construction is
examined. {\it Spectral \hyphenation{Darboux} Darboux coordinates} are
introduced on rational coadjoint orbits in $\wt{\frak{gl}}(r)^{+*}$, and
these are shown to generalize the hyperellipsoidal coordinates encountered
in the previous examples. Their relation to the usual algebro-geometric data,
consisting of linear flows of line bundles over the spectral curves, is given.
Applying the Liouville-Arnold integration technique, the Liouville generating
function is expressed in completely separated form as an abelian integral,
implying the Abel map linearization in the general case.
\medskip
\noindent {\bf Keywords.} Integrable systems, Liouville-Arnold integration,
loop algebras, isospectral flow, spectral Darboux coordinates, Abel map
linearization.

\newpage
\noindent{\sectionfont 1 Background Material and Examples}
\medskip \noindent
In this first section, we shall examine several examples of integrable
Hamiltonian systems that may be represented by isospectral flows on
coadjoint orbits of loop algebras. In each case, the flow may be linearized
through the classical Liouville-Arnold integration technique. An explicit
linearization of the flows can be made in terms of abelian integrals
associated to an invariant spectral curve associated to the system. A key
element in the integration is the fact that a complete separation of variables
occurs within a suitably defined coordinate system - essentially,
hyperellipsoidal coordinates, or some generalization thereof. A general theory
will then be developed in subsequent sections, based  essentially on  moment
map embeddings of finite dimensional symplectic vector  spaces, or Hamiltonian
quotients  thereof, into the dual space of certain loop algebras, the image
consisting of orbits whose elements are rational in the loop parameter.

  The origins of this approach may be found in the works of Moser {\bf [Mo]} on
integrable systems on quadrics,  Adler and van Moerbeke {\bf [AvM]} on
linearization of isospectral flows in loop algebras and the general
algebro-geometric integration techniques of Dubrovin, Krichever and Novikov
{\bf [KN], [Du]}. The theory of moment map embeddings in loop algebras is
developed in {\bf [AHP], [AHH1]}. Its relation to algebro-geometric
integration techniques is described in {\bf [AHH2]}, and the use of ``spectral
Darboux coordinates'' in the general Liouville-Arnold integration method in
loop algebras is developed in {\bf [AHH3]}. Some detailed examples and earlier
overviews of this approach may be found in {\bf [AHH4], [AHH5]}. The proofs of
the theorems cited here may be found in {\bf [AHP], [AHH1-AHH3]}.
\bigskip
 \noindent{\subsectionfont 1.1.\ The Neumann Oscillator  \hfill}
\medskip \noindent
We begin with the Neumann oscillator system ({\bf [N], [Mo]}), which consists
of a  point particle confined to a sphere in $\Bbb R^n$, subject to harmonic
oscillator forces.
The phase space is identifiable either with the cotangent bundle or the
tangent bundle (the equivalence being via the metric):
$$
M = T^*S^{n-1} =\{(\bold x,\bold y) \in \Bbb{R}^n \times \Bbb{R}^n \ \vert\
\bold x^T \bold x=1,\
\bold y^T \bold x=0\}
 \ss  \Bbb{R}^n \times \Bbb{R}^n, \tag{1.1}
$$
where $\bold x$ represents position and $\bold y$ momentum.
The Hamiltonian is
$$
H(\bold x, \bold y)={1\over2}[\bold y^T\bold y + \bold x^T A \bold x],
\tag{1.2}
$$
where $A$ is the diagonal $n \times n$ matrix
$$
 A=\text{diag}(\a_1, \dots \a_n) \in M^{n \times n},  \tag{1.3}
$$
with distinct eigenvalues $\{\a_i\}_{i=1, \dots n}$ determining the
oscillator constants.

   Equivalently, we may choose the Hamiltonian as:
$$
\phi(\bold x, \bold y)
= {1\over 2}[(\bold x^T \bold x)(\bold y^T \bold y) + \bold x^T A \bold x
- (\bold x^T \bold y)^2],  \tag{1.4}
$$
in view of the  constraints
$$
\bold x^T \bold x=1, \quad\bold y^T \bold x=0.\tag{1.5}
$$
The unconstrained equations of motion for the Hamiltonian $\phi$ are:
$$
\align
{d\bold x\over dt} & = (\bold x^T \bold x) \bold y -
(\bold x^T \bold y) \bold x  \tag{1.6a} \\
{d\bold y\over dt} & = - (\bold y^T \bold y) \bold x - A \bold x +
(\bold x^T \bold y) \bold y.  \tag{1.6b}
\endalign
$$
Since
$$
\{\phi, \bold x^T \bold x\}=0, \tag{1.7}
$$
it is convenient to interpret the relation
$$
n(\bold x):=\bold x^T \bold x  =1 ,  \tag{1.8}
$$
alone as a first class constraint. The function $n(\bold x)$ generates the
flow
$$
(\bold x, \bold y ) \lmt (\bold x, \bold y + t \bold x),  \tag{1.9}
$$
and is invariant under the $\phi$--flow.
We may then apply Marsden-Weinstein reduction, quotienting by the flow (1.9).
The reduced manifold is identified with a
section of the orbits under (1.9) defined by the other constraint:
$$
\bold
y^T \bold x =0.\tag{1.10}
$$
The integral curves for the constrained system are determined from those for
the unconstrained system by othogonal projection:
 $$
\align
(\bold x(t), \bold y(t))_{free} &\lmt
(\wh{\bold x}(t), \wh{\bold y}(t))_{constr.} \\
&:= \left( (\bold x(t), \bold y(t)-
\left({\bold x^T(t) \bold y(t)\over \bold x^T(t)\bold
x(t)}\right)\bold x(t) \right). \tag{1.11}
\endalign
$$
This lifts the projected flow on the Marsden-Weinstein reduced space
$n^{-1}(1)/\Bbb R$ to one that is tangential to the section defined by
eq.~(1.10).

     Assuming the constants $\a_i$ determining the oscillator strengths are
distinct, the integration proceeds (cf\. {\bf [Mo]}) by introducing the
Devaney-Uhlenbeck commuting integrals
$$
I_i := \sum_{j=1, j\neq i}^n {(x_iy_j - y_i x_j)^2\over
\a_i - \a_j} +  x_i^2,\tag{1.12}
$$
which satisfy
$$
\align
 \sum_{i=1}^n I_i &= \bold x^T \bold x  \tag{1.13a}\\
\sum_{i=1}^n \a_i I_i & = 2 \phi . \tag{1.13b}
\endalign
$$
Define
 the degree $n-1$ polynomial $\PP(\l)$ by
$$
{\Cal P(\l) \over a(\l)}
:= -{1\over 4} \sum_{i=1}^{n} {I_i \over \l - \a_i},\tag{1.14}
$$
where
$$
\align
 a(\l) &:= \prod_{i=1}^n(\l - \a_i), \\
\Cal P(\l)& = P_{n-1}\l^{n-1}+ P_{n-2} \l^{n-2} + \dots + P_0 .\tag{1.15}
\endalign
$$
Then on $T^*S^{n-1}$,
$$
\align
P_{n-1}&= -{1\over4} \bold x^T \bold x = -{1\over 4},\\
 P_{n-2}&= {1\over2} \phi .\tag{1.16}
\endalign
$$
An equivalent set of commuting integrals consists of the coefficients of
the polynomial $\{P_0, \dots, P_{n-2}\}$.
The  Liouville-Arnold tori  $\bold T$  are the leaves of the Lagrangian
foliation defined by the level sets:
$$
P_i=C_i. \tag{1.17}
$$

We now proceed to the linearization of the flows through the Liouville-Arnold
method. First, introduce hyperellipsoidal coordinates
$\{\l_\m\}_{\m=1, \dots n-1}$ and their conjugate momenta,
$\{\z_\m\}_{\m=1,\dots n-1}$, which are defined by:
$$
\align
\sum_{i=1}^n {x_i^2\over \l -\a_i} &=
{\prod_{\mu=1}^{n-1}(\l - \l_\mu)\over a(\l)} \tag{1.18a}\\
\zeta_\mu = {1\over 2} \sum_{i=1}^n{x_iy_i\over\l_\mu - \a_i}
&=\sqrt{{\Cal P(\l_\mu) \over a(\l_\mu)}}.  \tag{1.18b}
\endalign
$$
In terms of these, the canonical $1-$form is:
$$
\th=\sum_{i=1}^{n}y_idx_i \vert_{T^*S^{n-1}}=
\sum_{\mu=1}^{n-1}\zeta_\mu d\l_\mu. \tag{1.19}
$$
Restricting this to $\bold T$ determines the differential of the Liouville
generating function $S$:
 $$
\sum_{\mu=1}^{n-1} \zeta_\mu d\l_\mu \vert_{P_i=cst.} = dS=
\sum_{\mu=1}^{n-1}\sqrt{{\Cal P(\l_{\mu})\over a(\l_{\mu})}}d\l_{\mu},
\tag{1.20}
$$
which, upon integration, gives
$$
S=\sum_{\mu=1}^{n-1}\int_0^{\l_\mu}\sqrt{{\Cal P(\l)\over a(\l)}}d\l .
\tag{1.21}
$$
This is seen to be an abelian integral on the (generically) genus
$g=n-1$ hyperelliptic curve $\Cal C$
defined by:
$$
z^2 + a(\l)\PP(\l)=0.   \tag{1.22}
$$
The linearizing coordinates conjugate to the invariants $P_j$ are then:
$$
Q_j:= {\di S\over \di P_j}
= {1\over 2} \sum_{\mu=1}^{n-1}\int_0^{\l_\mu}{\l^j d\l
\over \sqrt{a(\l)\Cal P(\l)}} = b_jt , \tag{1.23}
$$
where, for $\phi = 2 P_{n-2}$,
$$
b_{n-2} = 2, \qquad b_j=0, \quad j < n-2  .  \tag{1.24}
$$

The map:
$$
(\l_1, \dots \l_{n-1}) \lra (Q_1, \dots Q_{n-1})  \tag{1.25}
$$
defined by eq.~(1.23) is, up to normalization, the Abel map from the symmetric
product $S^{n-1} \CC$ to the Jacobi variety $\JJ(\CC)$ of $\CC$:
$$
\bold A :S^{n-1}\Cal C \lmt \Cal J(\Cal C) \sim \Bbb C^{n-1}/\G, \tag{1.26}
$$
where $\G=$ is the  period lattice.

 We now turn to the interpretation of such systems as isospectral flows in a
loop algebra (cf. {\bf [AHP], [AHH3], [AHH4]}). Let
$$
\Cal N(\l)=\l Y+\Cal N_0(\l),  \tag{1.27}
$$
where
$$
\align
\Cal N_0 (\l) & :=
 {\l \over 2} \pmatrix
-\sum_{i=1}^n{x_iy_i \over \l-\a_i} &
- \sum_{i=1}^n{y_i^2 \over \l-\a_i}\\
 \ \sum_{i=1}^n{x_i^2 \over \l-\a_i} &
\ \sum_{i=1}^n{x_i y_i \over \l-\a_i}
\endpmatrix  \in \wt{\frak{sl}}(2)^{+*} \tag{1.28a}\\
 Y & :=
\pmatrix
0 & -{1\over 2}\\
0 & \ 0
\endpmatrix . \tag{1.28b}
\endalign
$$
We define a map $\wt J_A:\Bbb R^n \times \Bbb R^n \lra \wt{\frak{sl}}(2)^{+*}$
to the dual space of the positive half of the loop algebra $\wt{\frak{sl}}(2)$,
  relative to the standard splitting
$\wt{\frak{sl}}(2)=\wt{\frak{sl}}(2)^{+*} + \wt{\frak{sl}}(2)^{-*}$
into holomorhic parts inside and
outside a  circle $S^1$ in the complex $\l$--plane, containing the
$\a_i$'s in its interior region:
 $$
 \wt J_A:\ (\bold x, \bold y) \lmt \Cal N_0(\l) \in \wt{\frak{sl}}(2)^{+*} .
 \tag{1.29}
$$
This is a Poisson map with respect to the Lie poisson structure on
$\wt{\frak{sl}}(2)^{+*}$. The Hamiltonian $\phi$ is then given by restriction
of an elementary spectral invariant:
 $$
\phi(\bold x, \bold y) = - \text{tr}(\Cal N(\l)^2)_0
 := -{1\over 2 \pi i}\oint_{S^1}\text{tr}(\Cal N(\l)^2){d\l \over \l},
\tag{1.30}
$$
and all the other invariants $P_j$ may be similarly represented.
The equations of motion are seen to be equivalent to the Lax equation:
$$
{d\Cal N \over dt}= [\Cal B, \Cal N],  \tag{1.31}
$$
where
$$
\Cal B :=d\phi(\Cal N)_+=
\pmatrix
\ \bold x^T \bold y& \l + \bold y^T \bold y\\
- \bold x^T \bold x& -\bold x^T \bold y
\endpmatrix , \tag{1.32}
$$
and $(d\phi(\NN)_+$ signifies projection of the element
$d\phi(\NN) \in \wt{\frak{sl}}(2)$ to $\wt{\frak{sl}}(2)^{+}$.
 This is an example of the  {\it Adler-Kostant-Symes} (AKS) {\it theorem}
(to be explained more fully in Section 2). The spectral invariants (elements
of the AKS ring) are generated by the residues of the rational function
$$
\text{det}\left({\Cal N(\l)\over\l}\right)={\Cal P(\l) \over a(\l)}
= -{1\over 4} \sum_{i=1}^{n} {I_i \over \l - \a_i}. \tag{1.33}
$$

To see the relation with the standard algebro-geometric linearization methods
({\bf [Du], [KN], [AHH2-AHH3]}), we begin with the invariant spectral curve:
$$
\text{det}(\LL(\l) - z\Bbb I_2) = z^2 + a(\l)P(\l)=0,
\tag{1.34}
$$
where
$$
\LL(\l :={a(\l)\over \l}\NN(\l), \quad
z :=a(\l){\z\over\l}, \tag{1.35}
$$
and let
$$
\Cal M(\l,\z):= \left({\Cal N(\l)\over\l} -\z\Bbb I_2\right), \tag{1.36}
$$
where $\Bbb{I}_2$ is the $2 \times 2$ unit matrix.
Then the columns of the matrix $\wt{\Cal M}(\l,\z)$ of cofactors  of
${\Cal M(\l,\z)}$  are the eigenvectors of $\Cal N(\l)$ ($\z \l$ = eigenvalue)
on $\Cal C$:
$$
\wt{\Cal M}(\l,\z) =
  \pmatrix
\ {1 \over 2}\sum_{i=1}^n{x_iy_i \over \l-\a_i} -\z &
 {1 \over 2}\sum_{i=1}^n{y_i^2 \over \l-\a_i} +{1 \over 2}\\
  -{1 \over 2}\sum_{i=1}^n{x_i^2 \over \l-\a_i} &
-{1 \over 2}\sum_{i=1}^n{x_i y_i \over \l-\a_i} +\z
\endpmatrix . \tag{1.37}
$$
The hyperellipsoidal coordinates $\{\l_\mu,\z_\mu\}_{\mu=1, \dots ,n-1}$
define the finite part of the zero-divisor:
$$
\Cal D= \sum_{\mu=1}^{n-1} p(\l_\mu, \z_\mu) + p(\infty_1), \tag{1.38}
$$
i.e., the zeros of a section of the bundle $E \ra \CC$ dual to
eigenvector line bundle. This bundle can be shown generically to have degree
$n$, and thus to be an element of the Picard variety $E\in \text{Pic}^n$. The
Abel map then identifies the symmetric product  $S^{n-1}\Cal C$ with
the Jacobi variety $\JJ (\Cal C) \sim \text{Pic}^{0}$.
The linearity of the flow in $\text{Pic}^{n}$ follows from noting that
the Lax equation
$$
{d\Cal N \over dt}= [d\phi(\Cal N)_+, \Cal N], \tag{1.39}
$$
implies a linear exponential form for the transition function
$$
 \t(\l, z,t) = \text{exp}(\phi_z(\l, z)t) . \tag{1.40}
$$
\medskip
\noindent{\subsectionfont 1.2 \ Nonlinear Schr\"odinger (NLS) Equation \hfill}
\medskip \noindent
We now apply a similar analysis to the quasi-periodic solutions of the
cubically nonlinear Schr\"odinger equation (cf. {\bf [P1], [AHP], [AHH4]})
$$
u_{xx} + {\sqrt{-1}} u_t = 2|u|^2u.  \tag{1.41}
$$
Let
$$
 \Cal N(\lambda) :=
  \frac{\lambda}{2}
\pmatrix
 i\sum_{j=1}^n
\frac{|z_j|^2}{\lambda - \alpha_j}&-\sum_{j=1}^n
\frac{\overline{z}^2_j}{\lambda - \alpha_j} \\
-\sum_{j=1}^n\frac{z^2_j}{\lambda - \alpha_j}&-i
\sum_{j=1}^n\frac{|z_j|^2}{\lambda-\alpha_j}
\endpmatrix ,  \tag{1.42}
$$
and let $\o$ denote the standard symplectic form on $\Bbb C^n$:
$$
\omega =i\ d\overline{\bold z}^T\wedge d\bold{z}
= i \sum^n_{j=1}  d\overline{z}_j \wedge dz_j \tag{1.43}
$$
We define the Poisson map:
$$
\align
\wt{J}:  \Bbb C^n & \lra \wt{\frak{su}}(1,1)^{+*}\\
\wt{J}:\bold z & \lmt \Cal N(\lambda),   \tag{1.44}
\endalign
$$
and let
$$
{\Cal L}(\lambda) =\frac{a(\l)}{\l} \NN(\l)
= L_0\lambda^{n-1} + L_1 \lambda^{n-2} + \dots + L_{n-1}. \tag{1.45}
$$
The spectral curve is defined by the characteristic equation
$$
\align
\text{det}({\Cal L}(\l)- z I) &= z^2 + a(\lambda)\PP(\lambda) =0,  \\
\PP(\lambda) &:= P_{0} + P_{1}\lambda + \dots + P_{n-2}\lambda^{n-2},
\tag{1.46}
\endalign
$$
and has genus $g=n-2$.
Choosing the AKS Hamiltonians:
$$
\align
H_x& = \frac{1}{2}\left[ \frac{a(\lambda)}{\lambda^n} \lambda
\ \text{tr}(\Cal N(\lambda)^2)\right]_0
=-P_{2,n-3}  \tag{1.47a} \\
H_t &= \frac{1}{2} \left[\frac{a(\lambda)}{\lambda^n}\lambda^2
\ \text{tr}(\Cal N(\lambda)^2)\right]_0 = -P_{2,n-4}   \tag{1.47b}
\endalign
$$
gives the Lax equations
$$
\align
\frac{d}{dx} {\Cal L (\l)} &= [(dH_x)_+, {\Cal L(\l )}]
\tag{1.48b}
\\
\frac{d}{dt} {\Cal L(\l )} &= [(dH_t)_+,{\Cal L(\l )}], \tag{1.48b}
\endalign
$$
where
$$
\align
(dH_x)_+ &= \l L_0 + L_1   \tag{1.49a} \\
(dH_t)_+ &=
\l^2 L_0 + \l L_1 + L_2.   \tag{1.49b}
\endalign
$$
Choosing invariant constraints so that:
$$
\align
L_0 = \frac{i}{2} &
\pmatrix
1&0\\
0&-1
\endpmatrix  ,
\quad
L_1 = \left(\matrix  0 &{\overline{u}}\\u& 0 \endmatrix\right),
\\
L_2& = i\left(\matrix
|u|^2&{-\overline{u_x}}\\u_{x}&-|u|^2
\endmatrix\right) ,  \tag{1.50}
\endalign
$$
the compatibility conditions:
$$
{\di(dH_x)_+ \over \di t} - {\di (dH_t)_+ \over \di x} +
 [(dH_x)_+,(dH_t)_+]=0
 \tag{1.51}
$$
reduce to the NLS equation (1.41).
  To  obtain the quasi-periodic solutions, we introduce the
{\it spectral Darboux Coordinates}
$\{q, P, \l_\mu, \z_\mu\}_{\mu=1, \dots n-2}$,
analogous to the hyperellipsoidal coordinates above:
 $$
\align
 \sum_{i=1}^n \frac{z^2_i}{\lambda - \alpha_i}& = -{2 u
\prod^{n-2}_{\mu=1} (\lambda - \lambda_\mu)\over a(\lambda)}
\tag{1.52a}
\\
\z_\mu =
-{i\over 2} \sum_{i=1}^{n} {|z_i|^2\over \l_\mu - \a_i}
&=\sqrt{ -{\Cal P(\l_\mu)\over a(\l_\mu)}}
\tag{1.52b} \\
q := \ln(u),&   \qquad  P:=(L_0)_{22} , \tag{1.52c}
\endalign
$$
Then the symplectic form may be expressed as
$$
\omega = \sum_{\mu=1}^{n-2} d\l_{\mu} \wedge d \z_{\mu} + dq \wedge dP.
\tag{1.53}
$$

  As above, the spectral curve is invariant, and the coefficients of the
characteristic polynomial generate a complete set of commuting
integrals, so we may apply the Liouville-Arnold integration method.
The coordinates $(\l_\mu, \z_\mu)$ defined by (1.52a,b) again give the
finite part of the divisor of zeros of the eigenvectors of $\NN(\l)$, while
the remaining pair $(q,P)$ are determined by the additional spectral data at
$\l=\infty$. The Liouville generating function in this case becomes
$$
S=\sum_{\mu=1}^{n-1} \zeta_\mu d\l_\mu \vert_{P_i=cst.} + q P=
\sum_{\mu=1}^{n-1}\int_0^{\l_\mu}\sqrt{-{\Cal P(\l)\over a(\l)}}d\l
+ P\, \ln u \tag{1.54}
$$
and the linearizing coordinates conjugate to the $P_i$'s are
$$
Q_{i}  = {\di S \over \di P_i} =
\frac{1}{2} \sum^{n-2}_{\mu=1} \int^{\lambda_\mu}_0 \frac{\lambda^i
d\lambda}{{\sqrt{-a(\lambda)\PP_2(\lambda)}}}
 =b_i x+ c_i t,\qquad i = 0,\dots n-3 , \tag{1.55}
$$
which are abelian integrals of the first kind, and
$$
Q_{2,n-2}  ={ \di S \over \di P_i} =
{1\over 2} \sum^{n-2}_{\mu=1} \int^{\lambda_\mu}_0
\frac{\lambda^{n-2}d\lambda}{{\sqrt{-a(\lambda)\PP_2(\lambda)}}} -
\frac{\ln  u}{2 P_{2}} =b_{n-2}x +c_{n-2} t ,  \tag{1.56}
$$
which is an abelian integral of the third kind, the integrand having simple
poles at the two points $(\infty_1, \infty_2)$  over $\l=\infty$.
For the Hamiltonian  $H_x= -P_{n-3}$, we have $b_i= - \d_{i,n-3}$, $c_i=0$,
while for $H_t= -P_{n-4}$,  $b_i= 0$,  $c_i = - \d_{i,n-3}$.
An explicit formula for the function $u(x,t)$ may be obtained in terms of the
Riemann theta function $\th$ associated to the spectral curve by applying the
reciprocity theorem relating the two kinds of abelian integrals
(cf. {\bf [AHH4]}).
$$
u(x,t) =\text{exp}(q)
= \tilde{K}\ \text{exp}(bx + ct)
\frac{\theta(\bold A(\infty_2,p)  + t\bold U + x\bold V -
\bold K)}{\theta(\bold A(\infty_1,p)  + t\bold U + x\bold V - \bold K)},
\tag{1.57}
$$
where $\bold A: S^{n-2}\CC \lmt \Cal J(\Cal C) \sim
\Bbb C^{n-2}/\Gamma$ is the Abel map,  $\bold U, \bold V \in \Bbb
C^{n-2}$,  $b, c \in \Bbb C$ are  obtained from the vectors with components
$(b_i, c_i)$ on the RHS of eq.~(1.55), (1.56) by applying the linear
transformation that normalizes the abelian differentials in (1.55), and $\bold
K$ is the Riemann constant. \bigskip
\noindent{\subsectionfont 1.3 \  Sine-Gordon Equation \hfill}
\medskip \noindent
As a last example, consider the sine-Gordon equation
(cf. {\bf [HW], [P2], [AA]})
$$
\frac{\partial^2u}{\partial x^2}-\frac{\partial^2u}{\partial t^2}
= \sin(u). \tag{1.58}
$$
Let
$$
 \Cal N(\l) :=
 \l Y + \Cal N_0(\l) ,  \tag{1.59}
$$
where
$$
Y= \pmatrix
0 & -1\\
1 & 0
\endpmatrix, \quad
\Cal N_0(\lambda) :=2\l
\pmatrix
 b(\l) & c(\l) \\
-\bar{c}(\bar\l) & -b(\l)
\endpmatrix,  \tag{1.60}
  $$
with $b(\l),\,c(\l)$ given by
$$
\align
b(\l)&=\l\sum_{i=1}^p \left(\frac{-\varphi_i \bar\gamma_i}{\a_i^2-\l^2} +
\frac{\bar\varphi_i \gamma_i}{\bar\a_i^2-\l^2}\right) \tag{1.61a}\\
c(\l)&=\sum_{i=1}^p \left(\frac{\a_i\bar\gamma_i^2}{\a_i^2-\l^2}+
\frac{\bar{\a}_i\bar\varphi_i^2}{\bar\a_i^2-\l^2}\right), \tag{1.61b}
\endalign
$$
and $\varphi, \g \in \Bbb C^p$ complex vectors with components
$\{\varphi_i, \g_i\}_{i=1,\dots p}$. (Here $\a_i$, $\overline{\a}_i$,
$-\a_i$, $-\overline{\a}_i$ are assumed to be distinct.)

 Define
$$
a(\l):=\prod_{i=1}^p[(\l^2-\a_i^2)(\l^2-\bar\a_i^2)] . \tag{1.62}
$$
The symplectic form on $\Bbb C^p \times \Bbb C^p$ is given by:
$$
\o=
4\sum_{i=1}^p(d\gamma_i\wedge d\bar\varphi_i
+ d\bar\gamma_i \wedge d\varphi_i). \tag{1.63}
$$
Again, define a Poisson map:
$$
\align
\wt{J}:  \Bbb C^p \times \Bbb C^p  &\lra \wh{\frak{su}}(2)^{+*}\\
\wt{J}:(\varphi,\g) & \lmt \Cal N_0(\lambda),   \tag{1.64}
\endalign
$$
where the twisted loop algebra:
$$
\wh{\frak{su}}(2)^{+}\ss \wt{\frak{su}}(2)^{+}
\ss \wt{\frak{u}}(2)^{+} \ss \wt{\frak{gl}}(2,\Bbb C)^{+}  \tag{1.65}
$$
is defined as the fixed point set in $\wt{\frak{sl}}(2, \Bbb C)^{+}$  under
the  involutions;
$$
\align
\s_1: X(\l) &\lmt X^{\dag}(\bar{\l}) \\
\s_2: X(\l) & \lmt \pmatrix 1 & 0 \\ 0 & -1 \endpmatrix  X(-\l)
               \pmatrix 1 & 0 \\ 0 & -1 \endpmatrix.  \tag{1.66}
\endalign
$$
Let $n=2p$, and
$$
\align
{\Cal L}(\lambda)&:=\frac{a(\l)}{\l} \NN(\l) \tag{1.67a}\\
&= a(\l) Y + L_0\lambda^{2n-1} + L_1 \lambda^{2n-2} + \dots + L_{2n-1}.
\tag{1.67b}
\endalign
$$
The spectral curve is:
$$
\align
\det(\Cal L(\l)-zI)&=z^2+a(\l)P(\l)=0 \tag{1.68a}
  \\
P(\l)&=P_0+\l^2 P_1 + \dots + \l^{2n-2} P_{n-1} +\l^{2n}. \tag{1.68b}
\endalign
$$

   Choosing the AKS Hamiltonians:
$$
\align
H_\xi(X)& :=\frac 12 \text{tr}\left(\frac{a(\l)}{\l^2}(X(\l)+\l Y)^2\right)_0
 = -P_0 \\
H_\eta(X)& :=-\frac 12 \text{tr}\left(\frac{a(\l)}{\l^{2N}}(X(\l)+\l Y)^2
\right)_0 =P_{n-1}  \tag{1.69}
\endalign
$$
gives the Lax equations
$$
\align
\frac{d}{d\xi} {\Cal L (\l)} &= [A, {\Cal L(\l )}] \tag{1.70a}
\\
\frac{d}{d\eta} {\Cal L(\l )} &= [B,{\Cal L(\l )}] ,  \tag{1.70b}
\endalign
$$
where
$$
\align
-A=dH_\xi(\Cal N)_-&=\frac 1\l (L_{2n-1} + a(0)Y)\\
B = dH_\eta(\Cal N)_+&= L_0 + \l Y .  \tag{1.71}
\endalign
$$
Choosing the level set:
$$
P_0={1\over 16} \tag{1.72}
$$
 gives
$$
L_{2n-1}+a(0)Y={1\over 4} \pmatrix 0 & e^{i u} \\ -e^{-i u} & 0
\endpmatrix , \tag{1.73}
$$
where
$$
e^{i u}=a(0)(c(0)-1),  \tag{1.74}
$$
with $u$ real.
Then the compatibility conditions:
$$
{\di A \over \di \eta} - {\di B \over \di \xi} + [A,B]=0  \tag{1.75}
$$
reduce to the Sine-Gordon equation
$$
u_{xx}-u_{tt}=\sin u ,  \tag{1.76}
$$
where
$$
\xi= x+t, \qquad \eta= x-t . \tag{1.77}
$$

The quasi-periodic solutions  are obtained as in the previous example. The
unreduced spectral curve is defined by
 $$
 z^2 + \tilde{a}(\l)\tilde{P}(\l)=0,  \tag{1.78}
$$
and is again hyperelliptic, with genus $g=2n-1$.  Quotienting by the
involution
$$
(z,\l)\lmt (z,-\l)  \tag{1.79}
$$
gives a reduced curve with  genus $g=n-1$ defined by
$$
  \quad z^2 +\tilde{a}(E)\tilde{P}(E)=0, \tag{1.80}
$$
where
$$
\l^2:=E,\quad \tilde{a}(E);=a(\l),
\quad \tilde{P}(E) := P(\l). \tag{1.81}
$$
We also introduce the augmented curve, defined by
$$
  \tilde{z}^2 + E\tilde{a}(E)\tilde{P}(E)=0,  \tag{1.82}
$$
of genus $g=n$, where
$$
\tilde{z}:=z \l.  \tag{1.83}
$$

  The spectral Darboux coordinates are defined by
$$
\align
\tilde{c}(E_\mu)-1 &=0   \tag{1.84a}\\
\z_\mu\sqrt{E_\mu} &=2\tilde{b}(E_\mu), \qquad \mu=1, \dots n, \tag{1.84b}
\endalign
$$
where
$$
 \tilde{b}(E):=b(\l), \quad \tilde{c}(E):=c(\l). \tag{1.85}
$$
These again are interpreted as zeros of the sections of the dual to the
eigenvector line bundle associated to  $\NN(\l)$. The symplectic form is then
$$
\o=\sum_{\mu=1}^{n} dE_\mu\wedge d\zeta_\mu= -d\theta.\tag{1.86}
$$
The Liouville generating function is
$$
S(P_0, \dots,P_{n-1},E_1, \dots, E_n)=\sum_{\mu=1}^{n} \int_{E_0}^{E_\mu}
\sqrt{-\frac{\tilde{P}(E)}{E\tilde{a}(E)}}dE,  \tag{1.87}
$$
giving rise to the Abel map linearization:
$$
\align
Q_i &=\frac{\partial S}{\partial P_i}= -\frac 12
\sum_{\mu=1}^{N} \int_{E_0}^{E_\mu}
\frac{E^i}{\sqrt{-E\tilde{a}(E)\tilde{P}(E)}}dE  \tag{1.88a}\\
& =C_i+2\delta_{i,0}\xi-2\delta_{i,n-1}\eta,  \tag{1.88b}
\endalign
$$
which only involves abelian integrals of the first kind on the augmented
curve. In terms of theta functions, the solution may be expressed as
$$
\align
u&=-i\left(\sum_{\mu=1}^n \ln(-E_\mu)-\pi\right) \tag{1.89a}\\
&=-2i\,\ln\frac{\Theta(\bold A(p_0,0)-\bold U\eta- \bold V\xi-\bold K)}
{\Theta(\bold A(p_0,\infty)-\bold U\eta-\bold V\xi-\bold K)}+C, \tag{1.89b}
\endalign
$$
where $\bold{U}, \bold{V}$ are again obtained from the coefficients on the RHS
of (1.88) by applying the normalizing linear transformation to the abelian
differentials appearing in (1.88).
Full details for this case may be found in {\bf [HW]}.

    In the following two sections, a general approach to integrable systems
is developed, yielding all the above results as particular cases,
but allowing generalizations to more complex systems of higher rank.

\newpage
\noindent {\sectionfont 2 \ Moment Map Embeddings in Loop Algebras}
\bigskip
 \noindent{\subsectionfont 2.1 \  Phase Space and Loop Group Action  \hfill}
\medskip \noindent
We begin by defining the {\it generalized Moser space} (cf.~{\bf [AHP]}) to
be the symplectic vector space consisting of pairs $(F,G)$ of rectangular
$N\times r$ matrices:
$$
M=\{(F,G)\in M^{N,r} \times M^{N,r}\}  \tag{2.1}
$$
with symplectic form:
$$
\o= \text{tr }dF^T\wedge dG.          \tag{2.2}
$$

The loop algebra, denoted $\Lg$, consists of smooth maps from a circle
$S^1$, centred at the origin of the complex $\l$--plane, into $\frak{gl}(r)$,
$\frak{sl}(r)$,  or some subalgebra thereof.
$$
\align \Lg &= \wt{\frak {gl}}(r)  \quad (\text{or }
\wt{\frak {sl}}(r))\\ &= \{X(\l)\in {\frak gl}(r),\ \l \in S^1 \ss \Bbb{C}
 \cup \infty\}.   \tag{2.3}
\endalign
$$
 There is a natural splitting of $\Lg$
$$
\Lg =\Lgp+\Lgm,    \tag{2.4}
$$
as a vector space direct sum of the subalgebra $\Lgp$, consisting of elements
$X(\l)$ admitting a holomorphic extension to the interior of $S^1$, and
$\Lgm$, consisting of elements admitting a holomorphic extension to the
exterior, with normalization $X(\infty)=0$.  We identify $\Lg$ as a dense
subspace of its dual space $\Lg^*$ through the pairing
$$
 \align <\mu, X> &:=
\frac{1}{2\pi i} \oint_{S^1}\text{tr}\left(\mu(\l)X(\l)\right)\frac{d\l}{\l},
\tag{2.5}
\\
\mu \in \lgm &\ , \ X\in \Lgp.
\endalign
$$
Under this pairing, we have the identification
$$
(\Lgp)^* \sim \lgm, \quad (\Lgm)^* \sim \lgp,    \tag{2.6}
$$
where
$$
\Lg^* = \lgp + \lgm   \tag{2.7}
$$
similarly represents a decomposition of $\Lg^*$ into subspaces consisting
of elements holomorphic inside and outside $S^1$, but with the normalization
such that elements $\mu \in \lgp$ satisfy $\mu(0)=0$ (and hence the constant
loops are included on $\lgm$).

The loop  group $\LGp$ is similarly defined to consist of smooth maps
$g:S^1\ra Gl(r)$ which admit holomorphic extensions to the interior of $S^1$:
We define a Hamiltonian  action:
$$
\align
\LGp: M &\lra M \\
g(\l):(F,G) &\lra (F_g, G_g),  \tag{2.8}
\endalign
$$
where $(F_g, G_g)$ are determined by the decomposition
$$
\align
(A- \l I)^{-1}Fg^{-1}(\l) &= (A- \l I)^{-1}F_g  + F_{\text{hol}} \tag{2.9a}\\
g(\l)G^T(A- \l I)^{-1} &=G_g^T(A- \l I)^{-1} + G_{\text{hol}}.  \tag{2.9b}
\endalign
$$
Here  $A\in M^{N,N}$ is some fixed $N\times N$ matrix, with eigenvalues
in the interior of $S^1$, and $(F_{\text{hol}}, G_{\text{hol}})$ denote
the parts of the expressions on the left that are holomorphic in the interior
of $S^1$. This Hamiltonian action is generated by the
equivariant moment map:
$$
\align
\wt{J}^A:M &\lra \Lgp \sim \lgm\\
\wt{J}^A(F,G) &= \l G^T(A- \l I)^{-1}F,  \tag{2.10}
\endalign
$$
which is thus a Poisson map with respect to the Lie Poisson structure on
$\Lg^{+*}$. This map is not injective, its fibres being (generically) the
orbits of the subgroup $G_A:=\text{Stab}(A) \ss Gl(N)$ acting by conjugation
on $A$, and by the natural symplectic action on $M$.
$$
g:(F,G)\lra (gF, (g^{T})^{-1}), \quad g\in G_A \ss Gl(N)  \tag{2.11}
$$
The relevant phase space is therefore the quotient
$$
\align
M/G_A &\sim \Lg^*_A,   \tag{2.12}
\endalign
$$
which is identified with a finite dimensional Poisson subspace
$\Lg^*_A \ss \Lg^{+*}$ consisting of elements that are rational in the
loop parameter $\l$, with poles at the eigenvalues of $A$.
 \bigskip
\noindent{\subsectionfont 2.2 \  Simplest Case  \hfill}
\medskip \noindent
We now consider the simplest case, where $A$ is a diagonal matrix:
$$
A=\text{diag}(\a_1,\dots \a_1,\dots \a_k,\dots \a_k,\dots \a_n,\dots \a_n),
\tag{2.13}
$$
possibly with multiple eigenvalues $\{\a_i\}_{i=1, \dots n}$ of
multiplicity $k_i \leq r$,  all in the interior of $S^1$. The matrices
$(F,G)$ are decomposed accordingly as
$$
F=
\pmatrix
 F_1\\  \vdots  \\ F_i \\ \vdots  \\ F_n
\endpmatrix, \qquad
G=
\pmatrix
 G_1\\   \vdots \\ G_i \\ \vdots \\ G_n,
\endpmatrix,   \tag{2.14}
$$
where $(F_i, G_i)$ are the  $k_i \times r$ dimensional blocks corresponding
to the eigenvalues $\a_i$. For this case, $\NN_0(\l)$ has only simple poles:
$$
\NN_0(\l)=\wt{J}^A(F,G)=-\l \sum_{i=1}^n {G_i^TF_i \over \l - \a_i}
:= \l \sum_{i=1}^n {N_i \over \l-\a_i},  \tag{2.15}
$$
 with residue matrices $N_i$ generically of rank:
$$
\text{rk}(F_i)=\text{rk}(G_i) =k_i. \tag{2.16}
$$
The $Ad^*\LGp$--action for this case becomes:
$$
g(\l):\NN_0(\l)\lmt \l\sum_{i=1}^n{g(\a_i)N_ig^{-1}(\a_i) \over \l-\a_i},
\tag{2.17}
$$
which can be identified with the $Ad^*$--action of the direct product
group $Gl(r)\times \dots \times Gl(r)$ (n times) on  $[\frak{gl}(r)^*]^n$.

The fibres of the map $\wt{J}^A $  coincide with the orbits of
the block diagonal subgroup:
$$
G_A = Stab(A) = Gl(k_1) \times Gl(k_2) \times \dots \times Gl(k_n) \ss Gl(N),
\tag{2.18}
$$
under the action:
$$
(h_1, \dots h_i, \dots h_n) : (F_i,G_i) \lmt (h_iF_i, (h_i^T)^{-1}G_i)
\tag{2.19}
$$
This is also a Hamiltonian action, generated by the
``dual'' moment map:
$$
J_H(F,G):= (F_1G_1^T, \dots , \dots F_nG_n^T)
\in (\frak{gl}(k_1) \times \dots \times \frak{gl}(k_n))^*. \tag{2.20}
$$
The $Ad^*\LGp$ orbits  are then the level sets of the Casimir invariants:
$$
\text{tr}(F_iG_i^T)^l,\quad k=1, \dots n \quad l=1, \dots k_i. \tag{2.21}
$$

    More generally, the image of the moment map $\wt{J}^A$ is a Poisson
submanifold of $\Lg_A \ss \Lg^{+*}$ consisting of elements of the form
$$
\NN_0(\l) =\l \sum_{i=1}^n \sum_{l_i=1}^{p_i}
{N_i^{l_i} \over (\l - \a_i)^{l_i}},  \tag{2.22}
$$
where $p_i$ is the dimension of the largest Jordan block of $A$
corresponding to eigenvalue $\a_i$.
\bigskip
\noindent{\subsectionfont 2.3 \  Dynamics: Isopectral AKS Flows  \hfill}
\medskip
The Hamiltonian flows to be considered are those generated by elements
of the ring  of $\text{Ad}^*$--invariant functions  $\II(\Lg^*)$, restricted
to the translate $\l Y + \Lg_A$ of the subspace $\Lg_A$ by a
fixed element $ \l Y \in \Lg^{-*}$, where $Y\in \frak{gl}(r)$. (The latter
is an infinitesimal character for $\Lgm$, since it annihilates the
commutator of any pair of elements.) We denote the ring of elements so
obtained by
$$
\II_{\text{AKS}}^Y := \II(\Lg^*)\vert_{\l Y + \Lg_A},  \tag{2.23}
$$
and refer to it as the AKS (Adler-Kostant-Symes) ring.

Let
$$
\NN(\l) = \l Y + \NN_0(\l) \in  \l Y + \Lg_A.  \tag{2.24}
$$
We then have the fundamental theorem that underlies the integrability
of the resulting Hamiltonian systems, the Adler-Kostant-Symes theorem:
\proclaim {Theorem 2.1 (AKS)}\newline \noindent
1. \quad If $H \in \II_{\text{AKS}}^Y$,  Hamilton's equations are:
$$
X_H(\NN)={d\NN\over dt} = [(dH)_+,\NN]= - [(dH)_-,\NN]  \tag{2.25a}
$$
2. \quad If  $H_1, H_2 \in  \II_{\text{AKS}}^Y$,
$$
\{H_1,H_2\} = 0.  \tag{2.25b}
$$
\endproclaim
Thus, all the AKS flows commute, and are generated by isospectral
deformations determined by Lax equations of the form (2.25a).
In fact, it may be shown ({\bf [RS], [AHP], [AHH2]}) that on generic
coadjoint orbits of the form (2.22), these systems are completely integrable;
i.e., the elements of the Poisson commutative ring $\II_{\text{AKS}}^Y$
generate a Lagrangian foliation. Since the map (2.10) with image consisting
of elements of the form (2.22) is a Poisson map, and passes to the quotient
Poisson space $M/G_A$ to define an injective Poisson map, the same
results may be applied to the pullback $\wt{J}^A\circ H$ of any Hamiltonian
in the AKS ring $\II_{\text{AKS}}^Y$.
\proclaim {Corollary 2.2}
The results of Theorem 2.1 remain valid if the Hamiltonians $H_1, H_2$ are
replaced by $\wt{J}^A\circ H_1,\ \wt{J}^A\circ H_2$ on the space
$\l Y + \Lg_A$, identified with $M/G_A$ through $$
\NN(\l) = \l Y +\NN_0(\l) = \l Y + \tilde{J}_A(F,G). \tag{2.26}
$$
\endproclaim
\bigskip
\noindent{\subsectionfont 2.4 \  Reductions  \hfill}
\medskip \noindent
To obtain interesting examples, one usually must reduce the generic systems
described above in a manner that is consistent with the structure
of the dynamical equations. This generally consists of Hamiltonian
symmetry reductions involving either continuous or discrete symmetry groups.
(It may also involve symplectic, or more generally, Poisson constraints.)
We briefly summarize the  procedure for both types of symmetry reductions
below. The discrete  Hamiltonian reduction procedure is described in greater
detail in {\bf [HHM]}; the continuous, Marsden-Weinstein reduction is fairly
standard {\bf [AM]}. \medskip
\noindent{\subsubsectionfont 2.4.1 \  Discrete Reduction: \hfill}

   We consider discrete groups generated by elements $\s$ either of finite
order or generating compact orbits, which act on the space $M$
by symplectic diffeomorphisms, and as automorphisms of the loop algebra
$\Lgp$. Let
$$
\s:M \lra M  \tag{2.27}
$$
be such a symplectomorphism, and
$$
\s_{\frak{g}}:\Lgp \lra \Lgp  \tag{2.28a}
$$
the corresponding automorphism of $\Lgp$, with dual Poisson map
$$
\s_{\frak{g}}^*:\Lg^{+*} \lra \Lg^{+*}.  \tag{2.28b}
$$
 We assume that the moment map $\wt{J}^A$ intertwines these two actions, so
that the following diagram commutes
$$
\CD
M   @>                \s       >> M \\
@V\tilde{J}^AVV                @VV\tilde{J}^A V \\
\Lg^{+*}           @> \s_{\frak g}^* >> \Lg^{+*}
\endCD  \tag{2.29}
$$
It follows that $\wt{J}^A$ may be restricted to the fixed point sets
$$
M_{\s} \ss M, \qquad \tilde{\frak k}^+:= \Lgp_{\s} \ss \Lgp, \tag{2.30}
$$
and its restriction defines a moment map from the fixed point set
$M_\s$ to the dual space
$$
\tilde{\frak k}^{+*}:= \Lg_\s^{+*} \ss \Lg^{+*} \tag{2.31a}
$$
of the subalgebra
$$
\tilde{\frak k}^+:=  \Lgp_{\s} \ss \Lgp \tag{2.31b}
$$
 of fixed elements under $\s_{\frak{g}}$.
The results of Theorem 2.1 and Corollary 2.2 may then be applied on the
reduced spaces, provided the Hamiltonians in the ring $\II_{\text{AKS}}^Y$
are chosen to be invariant under the symmetry $\s_{\frak g}^*$.
\medskip
\noindent{\subsubsectionfont 2.4.2 \
 Continuous Hamiltonian Reduction \hfill}

  All the Hamiltonians in the ring $\II_{\text{AKS}}^Y$ are invariant
under the Hamiltonian group action given by conjugation of $\NN(\l)$
by $\l$-independent elements in the stability subgroup of $Y$:
$$
 G_Y :=\text{Stab}(Y) \ss Gl(r),
\qquad \frak{g}_Y:=\text{stab}(Y) \ss \frak{gl}(r). \tag{2.32}
$$
This action is generated by  a moment map $J_Y$, given by the
leading term $N_0$ of $\NN_0$, restricted to $\frak{g}_Y$
$$
J_Y :=N_0\vert _{\frak{g}_Y}, \tag{2.33}
$$
where
$$
\tilde{J}^A = \NN_0(\l) =N_0+N_1 \l^{-1}+ \dots. \tag{2.34}
$$
Since the elements of $\II^Y_{\text{AKS}}$ are $G_Y$ invariant, $J_Y$
is conserved under all the AKS flows. Fixing a level set
$$
J_Y = \mu_0 \in \frak{g}_Y^*, \tag{2.35}
$$
which we assume to be a regular value of $J_Y$,
and restricting to the coadjoint orbit $\OO_{\NN_0(\l)} \ss (\Lgp)^*$,
the reduced space is
$$
\OO_{\text{red}} := J_Y^{-1}(\m_0)/G_0,  \tag{2.36}
$$
where $G_0 \ss G_Y$ denotes the stability subgroup of $\m_0$.
The  reduced Hamiltonians $H_{\text{red}}$ on  $\OO_{\text{red}}$
are then given by
$$
H_{\text{red}}\circ \pi =H\vert_{J_Y^{-1}(\m_0)},
\quad H\in \II_{\text{AKS}}^Y, \tag{2.37}
$$
where
$$
\pi:J_Y^{-1}(\m_0) \lra J_Y^{-1}(\m_0)/G_0  \tag{2.38}
$$
denotes the projection map.
\bigskip
\noindent{\subsectionfont 2.5 \  Examples \hfill}
\medskip \noindent
We now indicate how the loop algebra formulation of examples like those
of Section 1 is obtained from the general scheme described above.
\medskip \noindent
{\subsubsectionfont
 2.5.1 Neumann Oscillator (and similar examples)
in $\wt{\frak{sl}}(2, \bold R)^{+*}$}

   Consider the case $r=2, k_i=1, n=N$. A discrete antilinear involution
gives the reality conditions:
$$
\a_i=\overline{\a}_i, \quad F=\overline{F},\quad G=\overline{G}  \tag{2.39}
$$
reducing $\wt{\frak{gl}}(2, \bold C)^*$ to  $\wt{\frak{gl}}(2, \bold R)^*$.
The stabilizer of $A= \text{diag }(\a_1, \dots , \a_n)$ consists
of the diagonal subgroup $G_A=\{\text{diag}(d_1, \dots ,d_n)\ss Gl(n)\}$
acting as
$$
\align
G_A: M&\lra M \\
(d_1, \dots d_n):\pmatrix \vdots\\F_i \\ \vdots\endpmatrix,
                   \pmatrix \vdots\\G_i \\ \vdots\endpmatrix
&\lmt \pmatrix \vdots\\d_i F_i \\ \vdots\endpmatrix,
                   \pmatrix \vdots\\d_i^{-1}G_i \\ \vdots\endpmatrix,
\tag{2.40}
\endalign
$$
where $F_i$ and $G_i$ are just $2$--component row vectors.
The moment map generating this action is just
$$
J_A(F,G) = (F_1 G_1^T, \dots , F_n G_n^T) \in \Bbb R^n, \tag{2.41}
$$
which coincides with the traces of the $2 \times 2$ residue matrices
$N_i$ in (2.15). Choosing the zero level set for these,
Marsden-Weinstein reduction is equivalent to the subgroup reduction
$\wt{\frak{gl}}(2)^+\sps \wt{\frak{sl}}(2)^+$. Choosing an appropriate
symplectic section gives the reduced parametrization:
$$
\align
F={1\over \sqrt{2}}(\bold x,\bold  y) &,
 \quad G={1\over \sqrt{2}}(\bold y, -\bold x) \tag{2.42}\\
\bold x, \bold y &\in \Bbb R^n.
\endalign
$$
The reduced symplectic form becomes
$$
\o =d\bold x^T \wedge d\bold y. \tag{2.43}
$$
The reduced moment map is
$$
\Cal N_0 (\l) = \wt{J}^A(F,G)=
 {\l \over 2} \pmatrix
-\sum_{i=1}^n{x_iy_i \over \l-\a_i} &
- \sum_{i=1}^n{y_i^2 \over \l-\a_i}\\
 \ \sum_{i=1}^n{x_i^2 \over \l-\a_i} &
\ \sum_{i=1}^n{x_i y_i \over \l-\a_i}
\endpmatrix.    \tag{2.44}
$$

Viewing this as defined on the symplectic vector space
$\Bbb R^n \times \Bbb R^n$, there is a residual fibration generated by the
finite group $(\Bbb Z_2)^n$ of reflections in the coordinate hyperplanes.
The $\wt{\frak{sl}}(2)^-$ character $\l Y$ may be expressed as:
$$
\l Y = \l \pmatrix a & b \\ c & -a \endpmatrix \in \wt{\frak{sl}}(2)^{-*},
 \tag{2.45}
$$
and the  resulting AKS flows involve isospectral deformations
of elements of the form:
$$
\Cal N (\l) =\pmatrix a &\ b\\c&  -a \endpmatrix + \Cal N_0 (\l). \tag{2.46}
$$

The  Hamiltonians  are chosen, as usual, from the AKS ring
$\II^Y_{\text{AKS}}(\wt{\frak{sl}}(2)^*)$.
In addition to the symmetry reductions already implemented, it is
possible to impose further symplectic constraints of the form
$$
f(\bold x, \bold y)=0, \quad g(\bold x, \bold y)=0, \quad \{f,\ g\} \neq 0,
\tag{2.47}
$$
and apply the standard methods for constrained systems. (Provided one of
these functions is in  the Poisson commutative ring
$\II^Y_{\text{AKS}}(\wt{\frak{sl}}(2)^*)$, the constrained Hamiltonians will
still commute.) The particular case of the above with
$$
\align
a&=0,\quad b=-{1\over 2}, \quad  c=0,  \tag{2.48a}\\
f&:= \bold x^T \bold x -1 =0, \quad g:=\bold y^T \bold x=0, \tag{2.48b}
\endalign
$$
and Hamiltonian (1.30), gives the Neumann oscillator system.
The invariant spectral curve is of the form
$$
\det (\Cal L(\l) -z \Bbb I)=z^2 +a(\l)\Cal P(\l)=0, \tag{2.49}
$$
where
$$
\Cal L(\l):= {a(\l)\over \l}\Cal N(\l),  \tag{2.50}
$$
and $\Cal P(\l)$ is generally a polynomial of degree $n-1$ or $n$, depending
on whether $a^2+bc$ vanishes or not.
\medskip
\noindent{\subsubsectionfont 2.5.2 \  The NLS Equation: Reduction to
$\wt{\frak{su}}(1,1)^+ $ \hfill }

   Again, we choose $r=2$, $k_i=1$, $n=N$. Similarly to the previous example,
the zero moment map reduction under
$$
G_A=Stab(A) =\Bbb C^\times \times \Bbb C^\times \times \dots
\times \Bbb C^\times \tag{2.51}
$$
is equivalent the subgroup  reduction
$\wt{\frak{gl}}(2, \Bbb C)^{+}\sps \wt{\frak{sl}}(2, \Bbb C)^{+}$.
Choosing a suitable symplectic section gives the parametrization
$$
\align
F={1\over \sqrt{2}}(\bold z, \bold w) &,
\quad G={1\over \sqrt{2}}(\bold w, -\bold z), \tag{2.52}\\
\bold z&,\ \bold w \in \Bbb C^n.
\endalign
$$

The further reality conditions
$$
\align
\a_i=\overline{\a}_i,&\quad \bold z=i\overline{\bold w} \\
F={1\over \sqrt{2}} (\bold z, i\overline{\bold z}) ,&
\quad G={1\over \sqrt{2}}(-i\overline{\bold z},\bold z) \tag{2.53}
\endalign
$$
give the discrete reduction
$\wt{\frak{sl}}(2, \Bbb C)^{+*}\sps \wt{\frak{su}}(1,1)^{+*}$
as the fixed point set under an antilinear involution.
On this real subspace, the symplectic form becomes
$$
\o =i d\overline{\bold z}^T \wedge d\bold z,  \tag{2.54}
$$
and the reduced moment map has the form.
$$
\Cal N_0 (\l) = \wt{J}^A(F,G)=
  \frac{\lambda}{2}
\pmatrix
 i\sum_{j=1}^n
\frac{|z_j|^2}{\lambda - \alpha_j}&-\sum_{j=1}^n
\frac{\overline{z}^2_j}{\lambda - \alpha_j} \\
-\sum_{j=1}^n\frac{z^2_j}{\lambda - \alpha_j}&-i
\sum_{j=1}^n\frac{|z_j|^2}{\lambda-\alpha_j}
\endpmatrix.  \tag{2.55}
$$
In this case, we choose the character $\l Y$ to vanish, so $\NN(\l)$
coincides with $\NN_0(\l)$.
The commuting flows are generated by the pair of commuting Hamiltonians:
$$
\align
H_x& = \frac{1}{2}\left[ \frac{a(\lambda)}{\lambda^n} \lambda
\ \text{tr}(\Cal N(\lambda)^2)\right]_0
\in \II_{\text{AKS}}^0(\wt{\frak{su}}(1,1)^{+*}) \tag{2.56a}  \\
H_t &= \frac{1}{2} \left[\frac{a(\lambda)}{\lambda^n}\lambda^2 \
 \text{tr}
(\Cal N(\lambda)^2)\right]_0
 \in \II_{\text{AKS}}^0(\wt{\frak{su}}(1,1)^{+*}).  \tag{2.56b}
\endalign
$$
Further invariant constraints are added to ensure
that the leading terms of the polynomial matrix (1.45) have the
form given in eq.~(1.49). Defining $\LL(\l)$ again as in eq. (2.50), the
resulting invariant spectral curve again has the form
$$
\align
\det (\Cal L(\l) -z \Bbb I)&\,=z^2 +a(\l)\Cal P(\l)=0,  \tag{2.57a}\\
\LL(\l) &:= {a(\l) \over \l}\NN(\l), \tag{2.57b}
\endalign
$$
where $\Cal P(\l)$ now is of degree $n-2$.
The Lax form (1.48a,b) of Hamilton's equations then follows from the AKS
theorem, and the compatibility condition (1.51) gives the NLS equation (1.41).
\medskip
\noindent
{\subsubsectionfont 2.5.3 \  Higher Rank Case.
 Two Component Coupled NLS System:
\newline \phantom{2.5.3 \ }Reduction to $\wt{\frak{su}}(1,2)^+$\hfill}

    As an illustration  of a system described by an algebra of higher rank,
we consider the case of the coupled two component cubically nonlinear
Schr\"odinger equation ({\it viz.} {\bf [AHP], [AHH2], [AHH3]}):
$$
\align
i u_t + u_{xx}&= 2u(|u|^2+|v|^2) \tag{2.58a}\\
i v_t + v_{xx}&= 2v(|u|^2+|v|^2). \tag{2.58b}
\endalign
$$

   In this case, we take $r=3$ and  $k_i=1$ for all $i$, so $n=N$.
The process of discrete and continuous symmetry reduction is applied
analogously to the preceding case, giving the sequence
$\wt{\frak{gl}}(3, \Bbb C)^{+*}\sps \wt{\frak{sl}}(3, \Bbb C)^{+*}
\sps \wt{\frak{su}}(1,2)^{+*}$. The reduced form of the resulting
pair of $n\times 3$ matrices $(F,G)$ is
$$
F=(\r, \eta, \z) , \quad G=(\r, -\eta, -\z), \tag{2.59}
$$
where $\eta,\ \zeta \in \Bbb C^n$ is a pair of complex $n$--component
column vectors and $\r \in \Bbb R^n$
is a real column vector with components
$$
\r_i =\sqrt{|\eta_i|^2 + |\z_i|^2}, \quad i=1, \dots n. \tag{2.60}
$$
The reduced symplectic form is
$$
\o =i (d\overline{\eta}^T \wedge d\eta + d\overline{\z}^T \wedge d\z),
\tag{2.61}
$$
so the components $(\eta_i, \zeta_i)_{i=1, \dots n}$ and their
complex conjugates provide a canonical coordinate system on $\OO_{\NN_0}$.
The reduced moment map has the form
$$
\align
\Cal N_0 (\l) &= \wt{J}^A(F,G) \\
&=  -i \l \sum_{j=1}^n {1\over \l -\a_i}
\pmatrix
\r_i^2  &\eta_i  \r_i & \z_i \r_i \\
-\overline{\eta}_i \r_i & -|\eta_i|^2 &-\overline{\eta}_i z_i \\
 -\overline{\z}_i \r_i & -\overline{\z}_i \eta_i& -|\z_i|^2
\endpmatrix, \tag{2.62}
\endalign
$$
so the coadjoint orbit may be identified with $\Bbb C^n \times \Bbb C^n$.
Again we choose the character $\l Y$ to vanish, so $\NN (\l)$ coincides with
$\NN_0(\l)$. As for the single component NLS equation, the commuting pair of
Hamiltonians for the two component CNLS case is chosen to be
$$
\align
H_x& = \frac{1}{2}\left[ \frac{a(\lambda)}{\lambda^n} \lambda
\ \text{tr}(\Cal N(\lambda)^2)\right]_0
\in \Cal I(\wt{\frak{su}}(1,2)^{+*})  \tag{2.63a} \\
H_t &= \frac{1}{2} \left[\frac{a(\lambda)}{\lambda^n}\lambda^2 \
 \text{tr}
(\Cal N(\lambda)^2)\right]_0
 \in \Cal I(\wt{\frak{su}}(1,2)^{+*}).   \tag{2.63b}
\endalign
$$
Defining, as before,
$$
{\Cal L}(\lambda) :=\frac{a(\l)}{\l} \NN_0(\l)
= L_0\lambda^{n-1} + L_1 \lambda^{n-2} + \dots + L_{n-1}, \tag{2.64}
$$
further invariant constraints must also be imposed, implying  that the
leading terms are of the form (cf.~{\bf[AHP]}):
$$
\align
L_0 &= {i\over 3}
\pmatrix 2 &\ 0 &\ 0\\ 0 &-1 & \ 0\\ 0 &\ 0 & -1 \endpmatrix  ,
\quad
L_1= \pmatrix 0 & \overline{u} &\overline{v} \\
 u & 0 & 0 \\v & 0 & 0 \endpmatrix, \\
L_2 &= i\pmatrix |u|^2+|v|^2 & -\overline{u}_x & -\overline{v}_x\\
u_x & -|u|^2 & -\overline{v} u\\
v_x &-\overline{u} v & -|v|^2 \endpmatrix . \tag{2.65}
\endalign
$$

  The Lax equations generated by the Hamiltonians (2.63a,b) have
the same form as eqs.~(1.48a,b), and the compatibility conditions
(1.51) are equivalent to the CNLS system (2.58a,b).
The invariant spectral curve in this case is a three sheeted
branched covering of $\Bbb P^1$ determined by an equation of the form
$$
\det (\Cal L(\l) -z \Bbb I)=z^3 +a(\l)z\Cal \PP(\l) + a(\l)^2 \QQ(\l)=0,
\tag{2.66}
$$
where $\PP(\l)$ and $\QQ(\l)$ are polynomials of degrees $n-2$ and $n-3$,
respectively.
\bigskip\bigskip
\noindent{\sectionfont 3. Spectral Darboux Coordinates and Liouville-Arnold
Integration}
\bigskip\noindent
In this section, the general method of linearization of AKS flows
in rational coadjoint orbits will be explained.  For simplicity, the
spectral properties of the matrix $A$ will be chosen as in Section 2.2, but
the method is equally valid in the more general case (see {\bf[AHH3]}). It
consists of two steps. First, a suitable generalization of the hyperellipsoidal
coordinates encountered in the examples of Section 1 is introduced, the
{\it spectral Darboux coordinates} (Theorem 3.2) associated to the invariant
 spectral curve $\CC$. These consist of families of canonical coordinates on
 coadjoint orbits $\OO_{\NN_0}$ of the type discussed in the preceding section,
 which are naturally associated to the spectral data of the matrix $\NN(\l)$.
  The second step consists of using a Liouville generating function $S$ to
compute the canonical transformation to coordinates conjugate to the spectral
invariants, in which the flow becomes linear. It turns out that for all
Hamiltonians in the AKS ring $\II^Y_{\text{AKS}}$ this generating function,
defined with respect to the natural isospectral Lagrangian foliation of
$\OO_{\NN_0}$,  may be expressed within the spectral Darboux coordinate system
in completely separated form (Theorem 3.3). It follows from the construction
that this transformation
 is given in terms of abelian integrals, showing that, in the general case,
the Abel map yields a linearization of the flows on the Jacobi variety
$\JJ(\CC)$ of the spectral curve. One thus arrives at the algebro-geometric
linearization
 results ({\it viz.} {\bf [Du], [KN], [AvM]}) entirely through classical
 Hamiltonian methods.
\bigskip
 \noindent{\subsectionfont 3.1 \ Phase Space and Group Actions \hfill}
\medskip \noindent
In the following, the phase space will initially be thought of as a coadjoint
orbit $\OO_{\NN_0}$ within the image of a moment map of the type introduced
in Section 2.
$$
\align
\wt{J}^A :M &\lra \Lg^{+*}  \tag{3.1a}\\
\wt{J}^A:(F,G) & \lmt \l G^T(A-\l\Bbb I_r)^{-1}F.  \tag{3.1b}
\endalign
$$
The image defines a finite dimensional Poisson submanifold
$$
\text{Im}(\wt{J}^A ):= \Lg_A \ss \Lg^{+*} \tag{3.2}
$$
which, in the simplest case, consists of elements of the form
$$
\Lg_A =\{\Cal N_0(\l) =\l \sum_{i=1}^n {N_i\over \l -\a_i}\}, \tag{3.3}
$$
where the ranks $\{k_i\}_{i=1, \dots n}$ of the residue matrices $N_i$
coincide with the multiplicities of the eigenvalues $\{\a_i\}_{i=1, \dots n}$
of the diagonal $N\times N$ matrix
$$
A= \text{diag}(\a_i, \dots,\a_i, \dots \a_n). \tag{3.4}
$$

The coadjoint action of the loop group ${\LGp}$ on $\Lg_A$ in this case
becomes equivalent to the coadjoint action of $(Gl(r))^n$
on $(\frak{gl}(r)^{*})^n$,
obtained by evaluating the group element $g(\l) \in \LGp$ at the poles
$\l=\a_i$:
$$
g: \{N_i\}\lmt \{g(\a_i)N_i g(\a_i)^{-1}\}.  \tag{3.5}
$$
It follows that the $\text{Ad}_{\LGp}^*$--orbits  are determined as
simultaneous level sets of the Casimir invariants of the separate residue
matrices $N_i$ under this action:
$$
\OO_{\NN_0} =\{ \l \sum_{i=1}^n {N_i\over \l -\a_i}\ \vert \
\text{tr} N_i^l = c_{il},\ l=1, \dots k_i\}. \tag{3.6}
$$

    The equations of motion induced by any element of the AKS ring
$\phi \in \II^Y_{\text{AKS}}$ have the Lax form:
$$
{d\NN (\l) \over dt} = \left[d\Phi(\NN)_+, \NN \right], \tag{3.7}
$$
where
$$
\phi =\Phi \vert_{\l Y + \Lg_A} \tag{3.8}
$$
is the restriction of the $Ad^*$--invariant element $\Phi \in \II(\Lg^*)$ to
the subspace consisting of elements of the form
$$
\NN(\l) = \l Y + \NN_0(\l), \quad \NN_0 \in \Lg_A. \tag{3.9}
$$
Define the $\frak{gl}(r)$--valued polynomial
$$
\Cal L(\l)  := {a(\l) \over \l} \Cal N(\l) = a(\l)Y + L_0 \l^{n-1} +
 \dots + L_{n-1}, \tag{3.10}
$$
where
$$
a(\l) :=\prod_{i=1}^n(\l -\a_i) \tag{3.11}
$$
is the minimal polynomial of $A$. This satisfies the equivalent Lax
equation
$$
{d\LL (\l) \over dt} = \left[d\Phi(\NN)_+, \LL \right]. \tag{3.12}
$$
The  invariant spectral curve $\CC_0$ is then determined by the characteristic
equation
$$
\text{det }(\LL(\l) - z \Bbb I_r) = 0 \tag{3.13}
$$
which, after suitable compactification, is viewed as an $r$--fold branched
cover of  $\Bbb{P}^1$, possibly having singularities over the points
$\l=\a_i$ due to the $r-k_i$ fold multiplicity of zero eigenvalues.
Other singularities could, of course, also occur, but for simplicity we again
place ourselves in a ``generic'' situation in order to be able to give the
main  results in as explicit form as possible, and therefore exclude this
possibility. The essential results remain valid without such simplifying
assumptions, but explicit formulae for the spectral polynomial, genus,
dimensions of orbits, and form of the abelian differentials must be modified
accordingly.

We assume henceforth, for simplicity, that the spectral curves $\CC_0$ have no
singularities other than those that arise over $\{\l=\a_i\}_{i=1, \dots n}$,
if $k_i < r-1$, due to the multiple zero eigenvalues of the residue matrices
$N_i $.  This implies in particular that the $N_i$'s, while not necessarily
diagonalizable, must lie on orbits that have the same dimensions as the
diagonalizable orbits whose nonzero eigenvalues are distinct; namely,
$k_i(2 r - k_i-1)$. We also assume that one of two conditions holds, which
excludes further singularities over $\l=\infty$:
\smallskip \noindent
 Case (a)\quad $Y=0$ and $L_0$  lies on a $Gl(r)$ orbit of the same dimension
($r(r-1)$) as those with simple spectrum.
\newline \noindent
Case (b)\quad $Y\neq 0$  and lies on a $Gl(r)$ orbit of the same dimension as
those with simple spectrum.
\smallskip \noindent
{\it Remark.}\  An effect of this assumption is to eliminate from consideration
 the example 2.5.3, which has singularities over $\l=\infty$.  However,
 this case may also be dealt with  ({\it viz.} {\bf [AHH3]}), by imposing a
further set of symplectic constraints defining a ``generic'' deformation class
of admissible spectral curves.

  In some cases, it is not the orbit $\OO_{\NN_0}$ itself that is the relevant
phase space,  but its reduction under the Hamiltonian action consisting of
conjugation by the stability subgroup $G_Y =\text{Stab }(Y) \ss Gl(r)$:
$$
g:\Cal N_0(\l) \lmt g \Cal N_0(\l) g^{-1}, \quad g\in G_Y. \tag{3.14}
$$
The corresponding moment map is just the leading term in
$\NN_0(\l)$:
$$
J(\Cal N_0):=L_0=\sum_{i=1}^n N_i, \tag{3.15}
$$
restricted to the subalgebra $\frak{g}_Y := \text{stab}(Y) \ss \frak{gl}(r)$.
Another case of interest, particularly when $Y=0$, consists of restricting to
a symplectic submanifold  $\OO^\SS_{\NN_0}\ss\OO_{\NN_0}$ determined by the
zero level set of the components of $L_0$ within the annihilator of a Cartan
subalgebra. (This is symplectic at regular elements $L_0$.)
For future reference, we list the various subcases of interest.
\smallskip\noindent
Case (a)\ $Y=0$. \
In this case, all the elements $\phi\in\Cal I\vert_{\Lg^{+*}}$ in the AKS ring
are invariant under the full $Gl(r)$ action (3.14), and all components of $L_0$
 are conserved. We may therefore reduce by the full group $Gl(r)$ or any of its
 subgroups. The two subcases of greatest interest are:
\smallskip \noindent  Case (a.1) \ {\it Complete reduction at a regular
point $L_0 =\mu_0 \in \frak{gl}(r)^*$.} The reduced manifold is then
$$
\OO_{\NN_0}^{\text{red}} = J^{-1}(\mu_0)/G_0, \tag{3.16}
$$
where $G_0 \ss Gl(r)$ is the stabilizer of $\mu_0$. The dimension of the
reduced orbit is:
$$
 \text{dim }\OO_{\NN_0}^{\text{red}} = \text{dim }\OO_{\NN_0}- (r-1)(r+2).
\tag{3.17}
$$
\noindent
Case (a.2)\ {\it Symplectic invariant manifold.} We take the zero level set
 of all components of $L_0$ in the annihilator of a Cartan subalgebra, (e.g.,
we choose $L_0$ to be diagonal). Denote this submanifold, which is symplectic
at all regular values of $L_0$, as:
$$
\Cal O_{\Cal N_0}^{\Cal S} :=\{\Cal N_0 \in O_{\Cal N_0} \vert
L_0 \in \Cal T \quad \text{(Cartan subalgebra)}\}.  \tag{3.18}
$$
Its dimension is
$$
\text{dim\ }\Cal O_{\Cal N_0}^{\Cal S} = \text{dim\ }O_{\Cal N_0} -r(r-1)
 = \text{dim\ }\Cal O_{\Cal N_0}^{\text{red}} +2(r-1).  \tag{3.19}
$$
\smallskip\noindent
Case (b)\ $Y\neq0$. \    In this case, the elements
$\phi\in\Cal I\vert_{\Lg^{+*}+\l Y}$ of the AKS ring are  only invariant under
the action of the stabilizer $G_Y =\text{Stab}(Y) \ss Gl(r)$. Two cases of
special interest arise:
\smallskip \noindent
Case (b.1) \ The full orbit $\OO_{\NN_0}$ (i.e., no reduction).
\smallskip \noindent
 Case (b.2) \quad The reduction of $\OO_{\NN_0}$ under the full stabilizer
 $G_Y\ss Gl(r)$ of a regular element $Y \in gl(r)$, taken at a value
$L_0\vert_{\frak{g}_Y} =\mu_0 \in \frak{g}_Y^*$. The group $G_Y$ is a maximal
 abelian subgroup with $r-1$ dimensional orbits and $G_0=G_Y$. The reduced
orbit is denoted
$$
\Cal O_{\Cal N_0}^{Y,\text{red}} = J\vert_{\frak{g}_Y}^{-1}(\mu_0)/ G_Y,
\tag{3.20}
$$
and has dimension
$$
\text{dim\ }\Cal O_{\Cal N_0}^{Y,\text{red}}
= \text{dim\ }\Cal O_{\Cal N_0} -2(r-1).  \tag{3.21}
$$
\bigskip
\noindent{\subsectionfont 3.2 \ Structure of the Spectral Curve  \hfill}
\smallskip \noindent
The affine part of the spectral curve $\CC_0$ is determined by the
characteristic equation (3.13). Taking into account the ranks of the residue
matrices $\{N_ i\}_{i=1, \dots n}$ in (3.3), we see that the characteristic
polynomial has the general form
$$
\align
\PP (\l,z) &=\text{det }(\LL(\l) - z \Bbb I_r)\\
&=(-z)^r + z^{r-1}\PP_1(\l) + \sum_{j=2}^rA_j(\l)\PP_j(\l)z^{r-j}, \tag{3.22}
\endalign
 $$
where
$$
A_j(\l) :=\prod_{i=1}^n(\l-\a_i)^{\text{max}(0, j-k_i)},
\quad \text{rank }\Cal L(\a_i)=k_i. \tag{3.23}
$$
This shows that near $\l \sim \infty$, we have
$$
z\sim \l^m,\quad
m := \cases n & \text{if } Y = 0  \ \text{(case (a))}\\
          n-1 & \text{if } Y \neq 0 \ \text{(case (b))}.\endcases  \tag{3.24}
$$
This suggests a compactification, not within $\Bbb {P}^2$, but rather in the
total space of a line bundle  over $\Bbb P^1 = U_0 \cup U_\infty$ (where
$U_0$, $ U_\infty$ denote the open disks obtained by deleting $\l=\infty$ and
$\l=0$, respectively),  with coordinate pairs $(\l, z)$ over $U_0$ and
$(\wt{\l}, \wt{z})$ over $U_\infty$ related by
$$
(\l, z) \lmt (\wt{\l}={1\over \l},\ \wt{z} ={z\over \l^m}) \quad
(\text{over } U_0\cap U_\infty). \tag{3.25}
$$
This is just the total space $\TT$ of the bundle $\OO(m)\ra \Bbb P^1$ whose
sheaf of sections consists of homogeneous functions of degree $m$.
The transformation (3.25) extends the affine curve $\CC_0$ defined by (3.13)
over $\l=\infty$, defining the compactification:
$$
\CC_0 \hra \CC \hra \TT. \tag{3.26}
$$

   The possible spectral curves so arising are branched $r$--sheeted covers of
$\Bbb P^1$, which within any given orbit of type (3.3), have $z$--values over
each $\l=\a_i$ that are fixed (being Casimir invariants of the coadjoint
action (3.5)). Of these, there are $k_i$ nonsingular points
$(\l=\a_i, z=\z_{ia})_{a=1, \dots k_i}$ corresponding to the nonzero
eigenvalues of $\LL(\a_i)$,  and the point $(\l=\a_i, z=0)$, which generically
is an $r-k_i$--fold ordinary singular point corresponding to the $r-k_i$--fold
zero eigenvalue. Figure 3.1 gives a visualization of  the spectral curves
$\CC$, embedded in $\TT$, as branched coverings of $\Bbb P^1$,
constrained to pass through these points.
\vskip 6.5 true cm
\indent
\special{picture branched.epsf}
\bigskip
\centerline{\bf Figure 3.1}
\bigskip
The detailed structure may be expressed more precisely by writing the form of
the characteristic polynomial $\PP(\l, z)$ for any spectral curve $\CC$ in a
neighborhood of a given curve $\CC_R$ as a perturbation of the characteristic
polynomial $\PP_R(\l, z)$ defining $\CC_R$.
\proclaim {Proposition 3.1 ({\bf [AHH3]})} In a neighborhood of the point
$\NN_R \in\OO_{\NN_0}$  with characteristic polynomial $\PP_{R}(\l,z)$, the
characteristic polynomial has the form:
$$
\PP(\l,z) \equiv \PP_{R}(\l,z) + a(\l)\sum_{j=2}^{r}a_j(\l)p_j(\l)z^{r-j},
\tag{3.27}
$$
where
$$
\align
a_j(\l) & = \prod_{i =1}^{n}(\l-\a_i)^{\text{max}(0, j-k_i-1)}  \tag{3.28}
\\
p_j(\l) & =: \sum_{a=0}^{\d_j}P_{ja} \l^a,  \tag{3.29}
\endalign
$$
and $\{p_j(\l)\}_{j=2,\dots r}$ are polynomials of degree:
$$
\align
\d_j &\equiv \text{deg}\ p_j(\l) = \cases d_j-j  \quad &\text{if} \quad Y=0
\\
 d_j  \quad & \text{if} \quad Y \neq 0
\endcases  \tag{3.30a}\\
d_j& \equiv \sum_{i=1}^{n}\text{min}(j-1, k_i ). \tag{3.30b}
\endalign
$$
 The number of independent spectral parameters $\{P_{ja}\}$,
($a=0,\dots \d_j + n-m - 1,\  j=2, \dots r$) is thus:
$$
d = \wt{g} +r - 1, \tag{3.31}
$$
where
$$
\wt{g} = \frac{1}{2}(r-1)(mr-2)- \frac{1}{2}\sum_{i=1}^{n}(r-k_i)(r-k_i -1)
\tag{3.32}
$$
is the genus of the (partially) desingularized spectral curve $\CC$ obtained
by separating branches at $\{\a_i, 0\}$. In a neighborhood of any generic
point on $\OO_{\NN_0}$, these spectral invariants are all independent.
\endproclaim

 The complete integrability of the systems under consideration on the various
coadjoint orbits $\OO_{\NN_0}$, and the reductions $\OO_{\NN_0}^{\text{red}}$,
$\OO_{\NN_0}^{Y,\text{red}}$ and symplectic submanifolds $\OO_{\NN_0}^{\SS}$
thereof, may be seen from the following Table of generic dimensions for the
various cases discussed above. (Note that the $P_{ia}$'s referred to do
not include the leading terms of the polynomials $p_j(\l)$ in eq.~(3.29).)
 Recall that the value of the genus $\wt{g}$ of the desingularized curve $\CC$
given in Proposition 3.1 depends on the value $m$, which is different in the
cases (a) and (b):
$$
\wt{g}=\wt{g}(m),
\quad m= \cases  n-1 & \text{for case (a)}\\
                 n & \text{for case (b).} \endcases  \tag{3.33}
$$
\bigskip
\centerline{\bf Table of Dimensions}
\nobreak
\bigskip
\centerline{$\matrix
    \text{Case} \quad  &\text{Dimension} &\quad \# P_{ia}\text{'s}
&\quad \#P_{i}\text{'s}\\
&&&\\
 \text{(a)} & \text{dim }\OO_{\NN_0}=2\wt{g}+(r-1)(r+2)
& \wt{g}& r-1\\  \text{(a.1)} & \text{dim }\OO_{\NN_0}^{\text{red}}=2\wt{g} &
\wt{g}& 0\\
  \text{(a.2)} & \text{dim }\OO_{\NN_0}^{\SS}=2(\wt{g}+r-1) & \wt{g}
& r-1\\
\text{(b.1)} & \text{dim }\OO_{\NN_0}=2(\wt{g}+r-1) & \wt{g} & r-1\\
\text{(b.2)} & \text{dim }\OO_{\NN_0}^{Y, \text{red}}=2\wt{g} &\wt{g}& 0 \\
\endmatrix $}
\bigskip

       Here, the notation $\{P_i\}_{i=2, \dots r}$ is used to denote the
components of $L_0$ evaluated on a basis of the relevent Cartan subalgebra
(not including the trivial central element, which is a Casimir invariant, and
hence constant on orbits). These are also elements of the ring
$\II^Y_{\text{AKS}}$, corresponding to the leading terms in the polynomials
$p_j(\l)$ in eq.~(3.29) for case (a) , and the next to leading terms for case
(b) (the leading terms being constant in the latter case), but they are
listed separately since, in cases (a.1) and (b.2), they are fixed through the
Marsden-Weinstein reduction procedure, and hence do not contribute to the
number of independent invariants on the reduced spaces. Moreover, these
elements enter again in Section 3.4 when defining the {\it spectral Darboux
coordinates} for these two cases.  It follows from the dimensions in the Table
and the independence of the commuting invariants that cases (a.1), (a.2),
(b.1) and (b.2) all give completely integrable systems.
\bigskip
 \noindent{\subsectionfont 3.3 \ Spectral Lagrangian Foliation \hfill}
\smallskip \noindent
The Lagrangian foliation given by fixing simultaneous level sets of the
invariants in the ring $\II^Y_{\text{AKS}}$ of spectral invariants (i.e.,
fixing the spectral curve $\CC$) is depicted below in Figure 3.2 for the
various cases  discussed above.
\medskip \noindent
\vbox{\hfil $\OO_{\NN_0}^{\SS}$ or $\OO_{\NN_0}$ ($\OO_{\NN_0}^{\text{red}}$
or  $\OO_{\NN_0}^{Y, \text{red}}$)\vfil}
\vskip -10pt
\centerline{
\vbox{\offinterlineskip
\halign{&\vrule#&
\strut\quad\hfil#\quad\cr
\omit {\Bl}&&\omit {\Bl}&&\omit {\Bl}&&\omit {\Bl}&&\omit
{\Bl}&\cr  \noalign{\hrule}\cr
& Liouville-Arnold&&{\Bl}&&{\Bl}&&{\Bl}&&{\Bl}&\cr
&\hfil Tori $\bold{T}\ \lra$&&{\Bl}&&{\Bl}&&{\Bl}&&{\Bl}&\cr
&(isospectral leaves)&&{\Bl}&&{\Bl}&&{\Bl}&&{\Bl}&\cr
&\hfil dim $\bold{T}=\hfil$&&{\Bl}&&{\Bl}&&{\Bl}&&{\Bl}&\cr
&$\wt{g}+r-1$ (or $\wt{g}$)&&{\Bl}&&{\Bl}&&{\Bl}&&{\Bl}&\cr
\noalign{\hrule}
\omit {\Bl}&&\omit {\Bl}&&\omit {\Bl}&&\omit {\Bl}&& \omit {\Bl}&\cr
\omit {\Bl}&&\omit {\Bl}&&\omit {\Bl}&&\omit {\Bl}&& \omit {\Bl}&\cr}
\hrule}}
\vskip -20 pt
\centerline{Admissable curves $\CC \ss \TT$}
\nobreak
\centerline{dim $\CC=\wt{g}+r-1$ (or $\wt{g}$)}
\nobreak
\smallskip \noindent
\centerline {\bf Figure 3.2}

\medskip
We may summarize the relevant spectral data associated to each element
$\NN \in \l Y +\LgA$  as follows:
\newline
$\bullet$ A {\it spectral curve} $\CC$ ($r$--fold branched cover of
$\Bbb P^1$) defined by the characteristic equation:
$$
\PP(\l,z) = \det(\LL(\l)-z \Bbb I_r)=0, \tag{3.34}
$$
(after suitable compactification and desingularization). The $r-1$ points
over  $\l=\infty$ are determined by the leading terms of the polynomials
$p_j(\l)$ of Proposition 3.1.
\newline
 $\bullet$ An {\it eigenvector subspace}:\ $[V(\l,z)] \ss \Bbb C^r$ at each
point in $\CC$ which, by our genericity assumptions, is one dimensional.

 Together, these determine an {\it eigenvector line bundle} $\check E \ra \CC$,
 and its corresponding dual bundle $ E \ra \CC$. The latter may be shown
({\it viz\. }{\bf[AHH2], [AHH3]}) to be generally of degree
$$
\text{deg }(E)= \wt{g}+r-1,   \tag{3.35}
$$
and hence, by the Riemann-Roch theorem, to have  an $r$--dimensional
space of sections. Conversely, it turns out that this data is sufficient to
reconstruct  the matrix $\LL(\l)$ (and hence $\NN(\l)$) {\it up to conjugation
by an element of $Gl(r)$}; that is, it is sufficient to determine the
projected point in the reduced orbit $\OO^{Y, \text{red}}_{\NN_0}$ or
 $\OO^{\text{red}}_{\NN_0}$, but not the element $\NN(\l)$ itself.

    In order to reconstruct the element $\NN(\l)$, it is necessary to add some
 further spectral data, consisting of a {\it framing} at $\l=\infty$; that is,
a basis of sections $\{\s_i \in H^0(\CC, E)\}_{i=1, \dots r}$ of the bundle
$E \ra \CC$, chosen to vanish, e.g. at all but one of the $r$ points
 $\{\infty_i\}_{i=1, \dots r}$ over $\l=\infty$ (for the case where the
Cartan subalgebra in question consists of the diagonal matrices).
$$
\s_i(\infty_j)=0, \quad i\neq j. \tag{3.36}
$$
This adds $r-1$ dimensions to the fibres, (since framings related by
\{$\wt{\s}_i = \k\s_i$\} are equivalent). Furthermore, the spectrum over
$\l=\infty$ in the class of admissible spectral curves must be left
undetermined, adding $r-1$ dimensions to the base space in Figure 3.2.

  More generally, it is insufficient to just consider line bundles, since this
excludes the possibility of degeneracy in the spectrum and further singular
points. The appropriate generalization consists of a coherent sheaf defined by
the exact sequence
$$
0 \lra \OO(-m)^{\oplus r} @> \LL^T(\l)-x\Bbb I>> \OO^{\oplus r} \lra E \lra 0,
\tag{3.37}
$$
where $\OO(-m)$ denotes the sheaf obtained by pulling back the corresponding
sheaf over $\Bbb P^1$ to $\TT$. In the case of bundles, the exact sequence
(3.37) just means that the dual  space to the space of eigenvectors  over the
spectral curve is given by the cokernel of the linear map defined by
$\LL^T(\l)-z\Bbb I_r$. A more complete discussion of the significance of this
construction may be  found in {\bf [AHH2], [AHH3]}.
\bigskip \noindent
{\subsectionfont 3.4 \ Spectral Darboux Coordinates \hfill}
\medskip \noindent
In this section we give a method for  constructing the appropriate Darboux
coordinates naturally associated to the spectral data discussed above, in
which the Hamiltonians in the spectral ring $\II^Y_{\text{AKS}}$ determine a
Liouville generating function in completely separated form.  First, we shall
give a purely computational description of these coordinates in terms of
simultaneous solutions of polynomial equations. The significance of this
prescription in terms of the eigenvector line bundles of the preceding section
will follow.

  Let
$$
M(\l,\z):= {\NN(\l)\over \l} - \z \Bbb I_r, \tag{3.38}
$$
and denote by $\wt{M}(\l,\z)$ the transpose of the matrix of
cofactors. Then, over the spectral curve defined by the characteristic
equation (3.34), the columns of $\wt{M}(\l,\z)$ are the eigenvectors of
$\LL(\l)$  (or $\NN(\l)$), and hence, generically, these are all proportional;
i.e. $\wt{M}(\l,\z)$ has rank $1$. Let $V_0 \in \Bbb C^r$ be a fixed vector,
and denote the solutions to the system of polynomial equations
$$
\wt{M}(\l,\z) V_0 =0,\quad
V_0\in \Bbb C^r \tag{3.39}
$$
as $ \{\l_\mu, \z_\mu\}_{\mu =1, \dots}$. Note that, due to the rank condition,
there really  are only two independent equations here, the other $r-2$
following as linear consequences.

   The significance of these equations in relation to the spectral data is
quite simple; they are the conditions that a section of the dual eigenvector
line bundle $E \ra \CC$ should vanish. The solutions give the zeros of the
components of the eigenvector determined by the vector $V_0$. As is well known
in algebraic geometry, giving the divisor of zeros of any section of a line
bundle amounts to giving the linear equivalence class of the bundle itself.
It follows, since the bundle $E \ra \CC$ is of degree $\wt{g}+r-1$, that
there will in general be $\wt{g}+ r-1$ zeros. However this is not necessarily
the number of solutions to (3.39), since some of the zeros may be over
$\l=\infty$. In fact, we may distinguish two cases of particular interest as
follows. In order to characterize the spectrum over $\l=\infty$, define
$$
\wt{\LL}(\l):=\LL(\l) / \l^m . \tag{3.40}
$$
\smallskip\noindent {\it Case (i).} \  $V_0$ is an eigenvector of
 $\wt{\LL}(\infty)$.  In this case, $r-1$ of the zeros are over $\l=\infty$
(the only point omitted over $\infty$ being the one corresponding to the
 eigenvalue of $V_0$.) Hence, there are only $\wt{g}$ finite solutions pairs
$\{\l_\mu, \z_\mu\}_{\mu=1, \dots \wt{g}}$, and these are generically
  independent, when viewed as  functions on the phase space $\OO_{\NN_0}^\SS$
(case (a)) or $\OO_{\NN_0}$ (case (b)). Moreover, they are invariant under the
 action of the reduction group for both cases, since this leaves the space
 $[V_0]$ invariant, and hence they project to functions on the reduced space
 $\OO_{\NN_0}^{\text{red}}$ (case (a)) or  $\OO_{\NN_0}^{Y,\text{red}}$ (case
 (b)). In view of the dimensions given in the Table of Dimensions, Section
3.2, the projected functions provide coordinate systems on the reduced
spaces.  On the prereduced spaces, we must supplement these with a further
$r-1$ pairs of coordinate functions, which we define as follows.  Choose a
basis where $L_0$ (case (a)) or $Y$ (case (b)) is diagonal, and
$V_0=(1, 0 \dots ,0)^T$. Then let
$$
\align
 q_i &:=\cases \text{ln}(L_1)_{i1} +
 {1\over 2} \sum_{j=2, j\neq i}^r \text{ln}(p_i -p_j) & \text{for case (a)}
\\
   \text{ln}(L_0)_{i1} & \text{for case (b)}
\endcases  \tag{3.41a}\\
P_i&:= (L_0)_{ii}, \quad i=2, \dots  r.  \tag{3.41b}
\endalign
$$
The pairs $\{q_i, P_i\}_{i=2, \dots r}$ provide the remaining coordinates
required.
\smallskip\noindent
{\it Case (ii).} \  $V_0$ is not an eigenvector of
$\wt{\LL}(\infty)$ and, furthermore,
$V_0 \notin \text{Im }(\wt{\LL}(\infty) -\wt{z}_i\Bbb I)$ for any eigenvalue
$\wt{z}_i(\infty)$ of $\wt{\LL}(\infty) $.  In this case, none of the zeros
are over $\l=\infty$, and there are generically $\wt{g} +r-1$ independent
solution pairs $\{\l_\mu, \z_\mu\}_{\mu=1, \dots \wt{g}+r-1}$ of eq.~(3.39).
These then provide a coordinate system on the prereduced space
$\OO_{\NN_0}^\SS$ (case (a)) or $\OO_{\NN_0}$ (case (b)).

We then have the following fundamental result.
\proclaim{Theorem 3.2} 1. \ If $V_0 \notin \text{Im }(\wt{\LL}(\infty)
-\wt{z}_i\Bbb I)$ for any eigenvalue $\wt{z}_i(\infty)$ of $\wt{\LL}(\infty)$,
the functions $\{\l_\mu, \z_\mu\}_{1, \dots \wt{g}+r-1}$ define a Darboux
coordinate system on $\OO_{\NN_0}^{\SS}$ (case (a) or $\OO_{\NN_0}$ (case (b)).
The orbital symplectic form is therefore:
$$
\o_{\text{orb}}= \sum_{\mu=1}^{\wt{g}+r-1} d\l_{\mu} \wedge d\z_{\mu}.
\tag{3.42a}
$$
2.\ If $V_0$ is an eigenvector of $L_0$ (case (a) or $Y$ (case (b)), the
functions $\{\l_\mu, \z_\mu\}_{1, \dots \wt{g}}$ project to Darboux
coordinates on the reduced spaces  $\OO_{\NN_0}^{\text{red}}$ (case (a)) or
$\OO_{\NN_0}^{Y,\text{red}}$ (case (b)), so the reduced orbital symplectic
form is:
$$
\o_{\text{red}}= \sum_{\mu=1}^{\wt{g}} d\l_{\mu} \wedge d\z_{\mu}.
\tag{3.42b}
$$
3. If $V_0$ is an eigenvector of $L_0$ (case (a)) or $Y$ (case (b)),
the functions \break $\{\l_\mu, \z_\mu, q_i, P_i\}_{\mu =1, \dots \wt{g},
i=2,\dots r}$ define a Darboux coordinate
system on $\OO_{\NN_0}^{\SS}$ (case (a) or $\OO_{\NN_0}$ (case (b)), so the
orbital symplectic form is:
$$
\o_{\text{orb}}= \sum_{\mu=1}^{\wt{g}} d\l_{\mu} \wedge d\z_{\mu}
+ \sum_{i=2}^r dq_i \wedge dP_i.  \tag{3.42c}
$$
\endproclaim

In the following section, we consider a number of elementary examples of the
above theorem. We shall see that the resulting {\it spectral Darboux
coordinates} do, indeed, generalize the hyperellipsoidal coordinates that were
encountered in the examples of Section 1.
\bigskip
\noindent{\subsectionfont 3.5 \ Examples \hfill}
\medskip \noindent
We begin by considering the simplest possible case; namely, where
 $\NN_0(\l)$ has only one pole,  at $\l=\a_1$, and $r=2$ or $3$. This just
corresponds to coadjoint orbits of the finite dimensional Lie algebras
 $\frak{sl}(2)$ and $\frak{sl}(3)$. Then we consider the case
 $\wt{\frak{sl}}(2,\bold R)^{+}$ for arbitrary $n$, with
$\text{rank}(N_i) =k_i =1$ for all $i=1, \dots n$. This reproduces the
hyperellipsoidal coordinates for the finite dimensional examples of Secs\. 1.1
and 2.5.1 (cf\. {\bf [Mo]}, such as the Neumann oscillator. Finally, we
consider the case $\wt{\frak{su}}(1,1)^+$, which provides the appropriate
complex coordinates  for the nonlinear Schr\"odinger equation, as discussed in
Secs\. 1.2 and 2.5.2.

\noindent
(a)\ {\it Single poles:} $n=1$
\newline \noindent
(a.1)\  Take $\frak{g}=\frak{sl}(2,\bold R)$, and (without
loss of generality), $\a_1=0$. Then  the dimension of a generic orbit is
dim~$\OO_{\NN_0} =2$.
We parametrize $\NN_0(\l)$ as follows:
$$
\NN_0(\l) = {\l N_1\over \l -\a_1} = N_1 := \pmatrix -a & r
              \\ u & a \endpmatrix,  \tag{3.43}
$$
and choose
$$
Y := \pmatrix 1 & 0 \\ 0 & -1 \endpmatrix, \quad V_0 = \pmatrix 1\\0
\endpmatrix.
\tag{3.44}
$$
The characteristic equation is then
$$
\text{det } (\LL(\l) - z \Bbb I_r)= z^2 - \l^2 - a^2 - u r =0.  \tag{3.45}
$$
In this case,  $V_0$ is an eigenvector of $Y$ and the genus of the spectral
curve is $\wt{g}=0$, so there are no $\{\l_\mu, \z_\mu\}$'s. The single pair
of spectral Darboux coordinates is thus
$$
q_2 = \text{ln }u, \quad P_2 = a.  \tag{3.46}
$$
It is easily verified that, relative to the Lie Poisson structure,
they satisfy
$$
 \quad \{q_2, P_2\} =1. \tag{3.47}
$$
(a.2)\  We consider the same orbit as in (a.1), but choose
$$
Y := \pmatrix 0 & 1 \\ 1 & 0 \endpmatrix.  \tag{3.48}
$$
In this case, $V_0$  is not an eigenvector of $Y$. The genus is still
$\wt{g}=0$, but the equation (3.39) now has a finite solution, giving the
Darboux  coordinate pair
$$
\l_1 = -u, \quad \z_1 =-{a \over u}. \tag{3.49}
$$
These are verified to also satisfy
$$
 \{\l_1, \z_1\}=0. \tag{3.50}
$$
(a.3)\ Take $\frak{g}=\frak{sl}(3, \bold R)$, and again, $\a_1=0$. Then the
dimension of a generic orbit with $n=1$ is dim $\OO_{\NN_0} =6$. We
parametrize $\NN_0(\l)$ as:
$$
\NN_0(\l) = N_1 := \pmatrix -a -b & r & s \\
               u & a & e \\
               v & f & b \endpmatrix, \tag{3.51}
$$
and choose
$$
Y = \pmatrix 0 & 0&0\\ 0 & 1 & 0 \\ 0 & 0 & -1  \endpmatrix,
 \quad V_0 = \pmatrix 1\\0\\0\endpmatrix.  \tag{3.52}
$$
Again, $V_0$ is an eigenvector of $Y$, but the spectral curve has genus
$\wt{g}=1$, and is realized as a $3$--fold branched cover of $\Bbb P^1$. We
therefore find one Darboux coordinate pair $(\l_1, \z_1)$, corresponding to a
finite zero of the eigenvector components, plus two further pairs,
$(q_2, P_2, q_3, P_3)$, corresponding to zeros over $\l=\infty$:
$$
\align
 \l_1 = {1\over 2} \left(b-a-{e v\over u} + {u f \over v}\right),&\quad
\z_1 = {u v  a + u v  b - e v^2 - f u^2 \over -u v  a + u v  b - e v ^2 +
 f u^2}
\\
q_2= \text{ln }u, \quad q_3= \text{ln }v,&\quad P_2 = a,\quad P_3 = b.
\tag{3.53}
\endalign
$$
Again, it is easily verified directly that these form a Darboux system,
with nonvanishing Lie Poisson brackets
$$
\{\l_1, \z_1\}=1,\quad \{q_2, P_2\} =1,\quad \{q_3, P_3\}=1.  \tag{3.54}
$$
\smallskip \noindent
(b) Now consider the case  $\Lg^+ =\wt{sl}(2, \bold R)^+$, with arbitrary $n$,
but ranks $k_i =1$ for all $i$, and hence $\text{det}(N_i)=0$ for all the
residue matrices $N_i$. For general $Y$, $\NN(\l)$ then has the form
$$
\Cal N (\l) = \l
\pmatrix
a & b \\
c & -a
\endpmatrix +
 {\l \over 2} \pmatrix
-\sum_{i=1}^n{x_iy_i \over \l-\a_i} &
- \sum_{i=1}^n{y_i^2 \over \l-\a_i}\\
 \ \sum_{i=1}^n{x_i^2 \over \l-\a_i} &
\ \sum_{i=1}^n{x_i y_i \over \l-\a_i}
\endpmatrix ,  \tag{3.55}
$$
where $\{x_i, y_i\}_{i=1, \dots n}$ form a Darboux system on the reduced
Moser space, which is identified with $\Bbb R^{2n}/(\Bbb Z_2)^N$. In this case,
the characteristic equation defining the invariant spectral curve $\CC$ is
$$
\text{det}(\LL(\l) - z\Bbb I_2) = z^2 + a(\l)P(\l)=0, \tag{3.56}
$$
where
$$
P(\l)= -(a^2+b c)\l^n + P_{n-1} \l^{n-1} + \dots.  \tag{3.57}
$$
In particular, this gives eq.~(1.34) for the case $a=c=0, \ b=-{1\over 2}$.
Thus, $\CC$ is hyperelliptic, a $2$--sheeted branched cover of $\Bbb P^1$,
with $2n-1$ or $2n$ finite branch points, depending on whether or not $a^2+bc$
vanishes. The genus is therefore generically $\wt{g}=n-1$. The dimension of
the coadjoint orbit is dim $\OO_{\NN_0} = 2n$. The matrix $\wt{\MM}(\l, \z)$ is
$$
\wt{M} =
 \pmatrix
-a+{1\over 2}\sum_{i=1}^n{x_iy_i \over \l-\a_i}-\z &
-b+{1\over 2}\sum_{i=1}^n{y_i^2 \over \l-\a_i}\\
-c -{1\over 2} \sum_{i=1}^n{x_i^2 \over \l-\a_i} &
a-{1\over 2} \sum_{i=1}^n{x_i y_i \over \l-\a_i} -\z
\endpmatrix . \tag{3.58}
$$
Taking
$$
V_0 =\pmatrix 1 \\ 0\endpmatrix,  \tag{3.59}
$$
if $c\neq 0$, $V_0$ is not an eigenvector of $Y$, so the full set of $n$
spectral Darboux coordinate pairs $\{\l_\mu, \z_\mu\}_{\mu=1, \dots n}$ are
given by:
$$
\align
& \sum_{i=1}^n {x_i^2 \over \l_\mu - \a_i} + 2c  =0,  \tag{3.60a}\\
&\z_\mu = -a + {1\over 2} \sum_{i=1}^n {x_i y_i \over \l_{\mu} - \a_i},
\tag{3.60b}\\
&\mu = 1, \dots n.
\endalign
$$
These are therefore hyperelliptic coodinates $\{\l_{\mu}\}$ and their
conjugate momenta $\{\z_\mu\}$. In the case $c=0$, $V_0$ is an
eigenvector of $Y$, and eqs.~(3.60a,b) are replaced by
$$
\align
 &{1\over 2}\sum_{i=1}^n {x_i^2 \over \l_\mu - \a_i}  =0,  \tag{3.61a}\\
&\z_\mu = -a + {1\over 2} \sum_{i=1}^n {x_i y_i \over \l_{\mu} - \a_i},
\tag{3.61b}\\
&\mu = 1, \dots n-1,
\endalign
$$
yielding only $n-1$ pairs of Darboux coordinates
$\{\l_\mu, \z_\mu\}_{\mu=1, \dots n-1}$, since one of the zeros of the
eigenvector components lies over $\l=\infty$. We must therefore complete the
system by defining the additional pair
$$
q:= \text{ln }({1\over2}\sum_{i=1}^n x_i^2), \quad  P:= {1\over 2}
\sum_{i=1}^n x_i y_i.  \tag{3.62}
$$
It is easily verified directly that
$$
\o = -d\theta, \tag{3.63}
$$
where
$$
\theta := \sum_{i=1}^n y_idx_i =
\cases \sum_{\mu =1}^n \z_\mu d\lambda_\mu &
\quad \text{if } c\neq 0  \\
 \sum_{\mu =1}^{n-1} \z_\mu d\lambda_\mu + Pdq &
\quad \text{if }c = 0 .
\endcases  \tag{3.64}
$$
\noindent
(c)\ {\it NLS Equation:} $\wt{\frak{su}}(1,1)^+$
\newline \noindent
Taking the orbit $\OO_{\NN_0}$ in $\wt{\frak{su}}(1,1)^{+*}$ as parametrized
in eqs.~(2.54), (2.55), with $Y=0$, the symplectic submanifold
$\OO^{\SS}_{\NN_0} \ss \OO_{\NN_0}$ is defined by the constraint
$$
\sum_{i=1}^n z_i^2 = 0. \tag{3.65}
$$
The spectral Darboux coordinates $\{q, P, \l_\mu, \z_\mu\}_{\mu =1, \dots n-1}$
are then given by eqs. (1.52a-c). It is easily verified in this case that
the orbital symplectic form restricted to $\OO^{\SS}_{\NN_0}$ is
$$
\o_{\text{orb}} = -d\theta, \tag{3.66}
$$
where
$$
\theta \vert_{\OO_{\NN_0}}=
 -i\sum_{j=1}^n \overline{z}_j d z_j \vert _{\OO_{\NN_0}}=
\sum_{\mu=1}^{n-2}\z_\mu d\l_\mu + Pdq, \tag{3.67}
$$
so $\{q, P, \l_\mu, \z_\mu\}_{\mu =1, \dots n-2}$ do, indeed, define a Darboux
coordinate system.

   A similar construction holds for the case of the sine-Gordon equation
(Section 1.3), where the relevant algebra is the twisted loop algebra
$\wh{\frak{su}}(2)^+$, obtained by a suitable combination of discrete and
continuous reductions. The orbits are parametrized by eqs.~(1.60), (1.61a,b),
and the relevant spectral Darboux coordinates determined by eqs.~(1.84a,b).
Details may be found in {\bf [HW]}.

   In the last section we explain how, in the general case, these spectral
Darboux coordinates lead directly to a linearization of the AKS flows through
the Liouville-Arnold integration procedure. In each case the relevant
linearizing map turns out to be the Abel map to the Jacobi variety of the
spectral curve. \bigskip
\noindent{\subsectionfont 3.6 \ Liouville-Arnold Integration \hfill}
\medskip \noindent
Using the spectral Darboux coordianates, we may define the local equivalent
of the ``canonical'' $1$--form
$$
\align
\theta &:= \sum_{\mu =1}^{\tilde{g}} \z_{\mu}d\l_{\mu} +
\sum_{i=2}^r P_idq_i  \tag{3.68a}\\
&\,= \sum_{\mu =1}^{\tilde{g}} {z_\mu\over a(\l_\mu)}d\l_{\mu} +
\sum_{i=2}^r P_idq_i, \tag{3.68b}
\endalign
$$
where
$$
z_\mu := a(\l_\mu)\z_\mu. \tag{3.69}
$$
(Note that for the examples given above, this actually {\it is} the canonical
$1$--form on $\Bbb R^{2n}/(\Bbb Z_2)^n$, or $\Bbb C^{2n}/(\Bbb Z_2)^n$, viewed
as the cotangent bundle of $\Bbb R^{n}/(\Bbb Z_2)^n$ and
$\Bbb C^{n}/(\Bbb Z_2)^n$, respectively.)
On the Liouville-Arnold torus $\bold{T}$,  defined by taking the simultaneous
level sets
$$
P_{ia}=C_{ia},\quad P_i = C_i \tag{3.70}
$$
of the spectral invariants, we have
$$
\theta\vert_{\bold T}=dS(\l_1, \dots \l_{\wt{g}}, q_2, \dots q_r, P_{ia},P_i),
\tag{3.71}
$$
where $S(\l_1, \dots, \l_{\wt{g}}, q_2, \dots, q_r, P_{ia}, P_i)$ is the
Liouville generating function to the canonical coordinates conjugate to the
invariants $(P_{ia}, P_i)$. Eq.~(3.71) can be integrated by viewing
$z=z(\l, P_{ia}, P_i)$ as a meromorphic function on the Riemann surface of the
spectral curve $\CC$.
$$
S(\l_{\mu}, q_i, P_{ia}, P_i) =
\sum_{\mu=1}^{\tilde{g}}\int_{\l_{\mu}^0}^{\l_{\mu}}
 \frac{z(\l, P_{ia}, P_j) }{a(\l)}d\l +
\sum_{i=2}^r q_iP_i, \tag{3.72}
$$
where $z_\mu=z_\mu(\l_\mu, P_{ia}, P_i)$ is essentially determined implicitly
by the spectral equation
$$
\PP(\l_\mu, z_\mu(\l_{\mu}, P_{ia}, P_i))=0.  \tag{3.73}
$$
The linearizing coordinates for AKS flows are then given, as usual, by
differentiation of $S$ with respect to the invariants:
$$
\align
Q_{ia} = \frac{\di S}{\di P_{ia}} &= \sum_{\mu=1}^{\tilde{g}}
\int_{\l_{\mu}^0}^{\l_{\mu}}\frac{1}{a(\l)}\frac{\di z}{\di P_{ia}}d\l
={\di h \over \di P_{ia}}t + Q_{ia,0}  \tag{3.74a}\\
Q_{i} = \frac{\di S}{\di P_{i}} &= \sum_{\mu=1}^{\tilde{g}}
\int_{\l_{\mu}^0}^{\l_{\mu}}\frac{1}{a(\l)}\frac{\di z}{\di P_{i}}d\l + q_i
={\di h \over \di P_{i}}t + Q_{i,0}, \tag{3.74b}
\endalign
$$
where, from the explicit structure (3.27) of the characteristic polynomial
given in Proposition 3.1, we obtain, by implicit differentiation, that the
integrands of eqs.~(3.74a,b) are of the form
$$
\align
\o_{ia} &:= \frac{1}{a(\l)}\frac{\di z}{\di P_{ia}}d\l
=\frac{a_i(\l)z^{r-i}\l^a}{\PP_z(\l,z)}d\l  \tag{3.75a}\\
\o_{i}&:=\frac{1}{a(\l)}\frac{\di z}{\di P_{i}}d\l
=-\sum_{j=2}^{r}\frac{R_{ij}a_j(\l)(-z)^{r-j}\l^{\d_j-\e}}{\PP_z(\l,z)}d\l,
\tag{3.75b}
\endalign
$$
where
$$
\align
R_{ij} &:=
\cases
& (P_1- P_i)\sum_{2\leq i_1 < i_2
\dots < i_{j-2} \neq i} P_{i_1} \dots P_{i_{j-2}}   \\
& \text{and} \quad \e=0\qquad  \text{for case (a) }\\
& (Y_1- Y_i)\sum_{2\leq i_1 < i_2
\dots < i_{j-2} \neq i} Y_{i_1} \dots Y_{i_{j-2}}  \\
& \text{and} \quad \e=1\qquad  \text{for case (b) }.\\
\endcases  \tag{3.76}
\endalign
 $$
The point to note is that the differentials $\{ \o_{ia}\}$, $\{\o_i\}$
 appearing in eqs. (3.75a,b) are, respectively, abelian differentials of the
first and third kinds on the Riemann surface defined by $\CC$,
the latter having their poles at the points $(\infty_1, \dots, \infty_r)$
over $\l=\infty$.
\proclaim
{Theorem 3.3 [\bf AHH3]} The $\wt {g}$  differentials $\{\o_{ia}\}_{i=1,
\dots \tilde{g}}$ in eq.~(3.75a) form a basis for the space
$H^0(\wt {\SS}, K_{\wt{\SS}}) $ of abelian differentials of the first kind
(where $K_{\wt {\SS}}$ denotes the canonical bundle). The linear flow
equation (3.74a) may therefore be expressed
as:
$$
 \bold{A}(\DD) = \bold{B} +\bold{U} t ,  \tag{3.77}
$$
where
 $\bold{A} :S^{\wt{g}} \CC \lra \Bbb C^{\wt{g}}/\Gamma$ is the Abel map, and
$\bold{B}, \bold{U} \in \Bbb{C}^{\wt{g}}$ are obtained by applying
the inverse of the $\wt{g} \times \wt{g}$ normalizing matrix $\bold{M}$,
with elements
 $$
\bold{M}_{\mu, (ia)} := \oint_{a_{\mu}} \o_{ia}, \tag{3.78}
$$
to the vectors $ \bold{C},\bold{H} \in \Bbb{C}^{\wt{g}}$ with components
$C_{ia}$  and $ -\frac{\di h}{\di P_{ia}}$, respectively (the pair $(ia)$
viewed as a single coordinate label in $ \Bbb{C}^{\wt{g}}$).
The $r-1$  differentials $\{\o_i\}_{i =2, \dots r}$ in  eq.~ (3.75b) are
abelian differentials of the third kind with simple poles at $\infty_i$ and
$\infty_1$, and residues $+1$ and $-1$, respectively.
 \endproclaim

  Comparing this general formula with the specific cases (1.23), (1.55),
(1.56), (1.88a,b), for the examples of Section 1, we see that this provides
the  generalization that was required, expressing all linearized AKS flows on
rational coadjoint orbits of $\wt{\frak{sl}}(r){+*}$ and its reductions
through the Abel map.

   It is possible, moreover, to invert the map expressing these flows, by
expressing any symmetric function of the coordinates
$\{\l_\mu, \z_\mu\}_{\mu =1, \dots \wt{g}}$ in
 terms of the Riemann theta function associated to the curve $\CC$. For
example, in view of eq.~(3.74b), the coordinates $\{q_i\}_{i=2, \dots r}$
 themselves are expressed as such symmetric functions through abelian
integrals of the third kind. Applying the reciprocity theorem relating
the two kinds of abelian integrals ({\it viz.} {\bf [AHH3]}), we obtain:
\proclaim {Corollary 3.4 [\bf AHH3]} For a suitable choice
of constants $\{e_i, f_i\}_{i=2, \dots r}$,  the coordinate functions
$\{q_i(t)\}$ satisfying eq.(3.74b)  are given by:
$$
q_i(t) = \ln \left[\frac{\theta(\bold{B}  +  t \bold{U} -\AB(\infty_i)-
\bold{K})} {\theta(\bold{B} + t \bold{U} -\AB(\infty_1) - \bold{K})}\right] +
e_i t +f_i,  \tag{3.79}
$$
where  $\bold{K} \in \Bbb{C}^{\wt{g}}$ is the Riemann constant.
  \endproclaim

   This generalizes the theta function formula (1.57) giving the solution of
 the NLS equation. Similar formulae exist e.g.,  for the sine-Gordon equation
 ({\bf [HW]}) and many other systems that can be cast in terms of commuting
AKS flows in rational coadjoint orbits of loop algebras.
Aside from technical complications resulting, e.g.,  from the reduction
 procedure  or the imposition of further symplectic constraints, or from
 the presence of further singularities in the spectral curve, the
 procedure is largely algorithmic. It provides a very general setting
 for the explicit application of the Liouville-Arnold integration procedure to
a wide class of known - and yet to be discovered - integrable Hamiltonian
systems. Moreover, the moment map embedding method makes it possible to treat
both the intrinsically finite dimensional systems, and those
 systems corresponding to finite dimensional sectors of integrable systems of
 PDE's (solitons, finite band solutions, etc.) on exactly  the same footing.
\bigskip \bigskip \noindent
{\it Acknowledgements.}\quad Most of the results described here were obtained
in collaboration with a number of colleagues and friends, whose
important contributions to this work it is a pleasure to acknowledge. My
thanks to M.~Adams, J.~Hurtubise, E.~Previato and M.-A.~Wisse for their
valuable input  and help over an extended period. I would also like
to thank G.~Helminck and the other organizers of the $8$th Scheveningen
conference for their very kind hospitality and for the cordial and
stimulating environment they helped to create.

\newpage
\centerline{\smc References}
\bigskip
{\smaller{
\item{\bf [AA]} Al'ber, S.J., Al'ber, M.S.,
``Hamiltonian Formalism for Nonlinear Schr\"odinger Equations and
Sine-Gordon Equations'', {\it J\. London Math\. Soc\. (2)}
{\bf 36}, 176--192 (1987).
\item {\bf [AM]} Abraham, R., Marsden, J.E.,
{\sl Foundations of Mechanics}, 2nd ed.,
Reading, MA, Benjamin Cummings, Ch. 4 (1978).
\item{\bf [AHH1]} Adams, M.R., Harnad, J. and Hurtubise, J.,
``Dual Moment Maps to Loop Algebras'', {\it Lett\. Math\. Phys\.} {\bf 20},
 294--308 (1990).
\item{\bf [AHH2]} Adams, M.R., Harnad, J. and Hurtubise, J., ``Isospectral
Hamiltonian Flows in Finite and Infinite Dimensions II.
Integration of Flows'',
 {\it Commun\. Math\. Phys\.} {\bf 134}, 555--585 (1990).
\item{\bf [AHH3]}  Adams, M.~R., Harnad,~J. and  Hurtubise,~J., ``Darboux
Coordinates and Liouville-Arnold Integration  in Loop Algebras,''
  {\it Commun\.  Math\.  Phys\./} (1993, in press).
\item{\bf [AHH4]} Adams, M.R., Harnad, J. and Hurtubise, J.,
``Liouville Generating Function for Isospectral Hamiltonian
Flow in Loop Algebras'', in:  {\it Integrable and
Superintergrable Systems}, ed. B. Kuperschmidt, World Scientific,
Singapore (1990).
\item{\bf [AHH5]} Adams, M.R., Harnad, J. and Hurtubise, J.,
``Coadjoint Orbits, Spectral Curves and Darboux Coordinates'',
in: {\it The Geometry of Hamiltonian Systems}, ed\. T. Ratiu,
Publ. MSRI Springer-Verlag, New York (1991); ``Integrable Hamiltonian Systems
on Rational Coadjoint Orbits of Loop Algebras'', in: {\it Hamiltonian Systems,
Transformation  Groups and Spectral Transform Methods},
ed. J. Harnad and J. Marsden, Publ. C.R.M., Montr\'{e}al  (1990).
\item{\bf [AHP]} Adams, M.R., Harnad, J. and Previato, E., ``Isospectral
Hamiltonian Flows in Finite and Infinite Dimensions
I. Generalised Moser Systems and Moment Maps into Loop Algebras'',  {\it
Commun\. Math\. Phys\.} {\bf 117}, 451--500 (1988).
 \item{\bf[AvM]} Adler, M.\ and van Moerbeke, P., ``Completely Integrable
Systems, Euclidean Lie Algebras, and Curves,''  {\it Adv.\  Math.\/}
{\bf 38}, 267--317 (1980); ``Linearization of Hamiltonian Systems, Jacobi
Varieties and Representation Theory,''
{\it ibid.} {\bf 38}, 318--379 (1980).
\item{\bf [Du]} Dubrovin, B.A., ``Theta Functions and Nonlinear Equations'',
 {\it Russ\. Math\. Surv\.} {\bf 36}, 11--92 (1981).
\item{\bf [KN]} Krichever, I.M.and  Novikov, S.P.,  ``Holomorphic Bundles over
Algebraic Curves and Nonlinear Equations'', {\it Russ\. Math.~ Surveys}
{\bf 32}, 53--79 (1980).
\item{\bf [HW]} Harnad, J. and Wisse, M.-A.  ``Isospectral Flow in Loop
Algebras and Quasiperiodic Solutions of the Sine-Gordon Equation'',
{\it J\. Math\. Phys\.} (1993, in press).
\item{\bf [Mo]}  Moser, J.,  ``Geometry of Quadrics and Spectral Theory", {\it
The Chern Symposium, Berkeley, June 1979}, 147--188, Springer,
 New York, (1980).
\item{\bf [N]} Neumann, C., ``De problemate quodam mechanico, quod ad
primam integralium ultraellipticorum classem revocatur'',
 {\it J. Reine Angew\. Math\.} {\bf 56}, 46--63 (1859).
\item{\bf [P1]} Previato, E., ``Hyperelliptic quasi-periodic and soliton
solutions of the nonlinear \break Schr\"odinger equation.''
{\it Duke Math\. J\.}
{\bf 52}, 329--377 (1985).
\item{\bf [P2]} Previato, E. ``A particle-system model of the
sine-Gordon hierarchy, Solitons and coherent structures'',
{\it Physica D} {\bf 18}, 312--314, (1986).
 \item{\bf [RS]} Reiman, A.G., and Semenov-Tian-Shansky,
M.A., ``Reduction of Hamiltonian systems, Affine Lie algebras and
Lax Equations I, II'', {\it Invent\. Math\.} {\bf 54}, 81--100 (1979);
 {\it ibid\.} {\bf 63}, 423--432 (1981).
\medskip}}
\vfil\eject
\enddocument